\documentclass[journal]{IEEEtran}
\ifCLASSINFOpdf

\else

\fi

\usepackage{mdframed}
\usepackage{xcolor}
\usepackage{CJK}
\usepackage{pifont}
\usepackage{mathrsfs}
\usepackage{algorithmic}
\usepackage{amsmath}
\usepackage{amssymb}
\usepackage{psfrag}
\usepackage{graphicx}
\usepackage{epstopdf}
\usepackage{multirow}
\usepackage{stfloats}
\usepackage{cite}
\usepackage{subcaption}
\usepackage{color}
\usepackage[normalem]{ulem}
\usepackage{arydshln}
\usepackage{enumerate}
\usepackage{bbm}
\usepackage{verbatim}
\usepackage{url}
\usepackage[hidelinks]{hyperref} 
\usepackage[T1]{fontenc}
\usepackage[utf8]{inputenc}

\usepackage{booktabs,tabularx,array}

\newtheorem{theorem}{\bf{Theorem}}[section]
\newtheorem{lemma}{\bf{Lemma}}[section]

\newcommand{\bm}[1]{\mbox{\boldmath{$#1$}}}

\def\BibTeX{{\rm B\kern-.05em{\sc i\kern-.025em b}\kern-.08em
    T\kern-.1667em\lower.7ex\hbox{E}\kern-.125emX}}
\usepackage{balance}

\newmdenv[
  backgroundcolor=white,
  linewidth=0.75pt,
  linecolor=black,
  leftline=true,
  topline=true,
  rightline=true,
  bottomline=true
]{mybox}
\usepackage[linesnumbered,ruled,vlined]{algorithm2e}

\begin{document}
\title{Beyond ISAC: Toward Integrated Heterogeneous Service Provisioning  via Elastic Multi-Dimensional Multiple Access}

\author{
Jie Chen, \IEEEmembership{Member, IEEE},   Xianbin Wang, \IEEEmembership{Fellow, IEEE}, and Dusit Niyato, \IEEEmembership{Fellow, IEEE}

\thanks{
This manuscript was submitted to IEEE Transactions on Communications on 9 November 2025, revised on 5 January 2026, and accepted on 6 January 2026.
The work of Jie Chen and Xianbin Wang was supported in part by the Natural Sciences and Engineering Research Council of Canada (NSERC) Discovery Program under Grant RGPIN-2024-05720 and in part by the Canada Research Chair Program under Grant CRC-2023-00336.
The work of Dusit Niyato  was supported by Seatrium New Energy Laboratory, Singapore Ministry of Education (MOE) Tier 1 (RT5/23 and RG24/24), and
 the Nanyang Technological University (NTU) Centre for Computational Technologies in Finance (NTU-CCTF).
The work in this paper was partially presented at IEEE International Conference on Communications 2025 \cite{chen2025icc}. (Corresponding author: Xianbin Wang.)

J. Chen and X. Wang are with the  Department of Electrical and Computer Engineering, Western University, London, ON N6A 5B9, Canada (e-mails: jiechen@ieee.org, xianbin.wang@uwo.ca). Dusit Niyato is with the College of Computing and Data Science, Nanyang Technological University, Singapore 639798 (email: dniyato@ntu.edu.sg).
 }
}

\maketitle

\begin{abstract}
Due to the growing diversity of vertical applications, current integrated sensing and communications (ISAC) technologies in wireless networks remain insufficient to support complex services beyond communications.
To this end, future networks are evolving toward an integrated heterogeneous service provisioning (IHSP) platform, which aims to integrate a broad range of heterogeneous services beyond the dual-function scope of ISAC.
Nevertheless, this trend intensifies the conflicts among concurrent heterogeneous services under constrained resource sharing.
In this paper, we overcome this resource constraint by the joint use of two novel elastic design strategies: compromised service value assessment and flexible multi-dimensional resource sharing.
Consequently, we propose a value-prioritized elastic multi-dimensional multiple access (MDMA) mechanism for IHSP.
First, we define the compromised Value-of-Service (VoS) metric by incorporating elastic parameters to characterize user-specific tolerance and compromise in response to various performance degradations under constrained resources.
This VoS metric serves as the foundation for prioritizing resource sharing among IHSP services with fairness among concurrent competing demands.
Next, we adapt the MDMA to elastically multiplex services using appropriate multiple access schemes across different resource domains.
This protocol leverages user-specific interference tolerances and cancellation capabilities across different domains to reduce resource-demanding conflicts and co-channel interference within the same domain.
Then, we maximize the system's VoS by jointly optimizing MDMA design and power allocation.
Since this problem is non-convex, we propose a monotonic optimization-aided dynamic programming (MODP) algorithm to obtain its optimal solution. Additionally, we develop the VoS-prioritized successive convex approximation (SCA) algorithm to efficiently find its suboptimal solution.
Finally, simulations are presented to validate the effectiveness of the proposed designs.
\end{abstract}

\IEEEpeerreviewmaketitle

\section{Introduction}

Future sixth-generation (6G) networks are expected to support both sensing and communication functionalities to meet the diverse demands of emerging applications \cite{liu2022integrated}.
To fulfill these dual requirements under constrained spectral and hardware resources, integrated sensing and communication (ISAC) has been recognized as a foundational technology for 6G, as outlined in the International Mobile Telecommunications (IMT)-2030 framework~\cite{series2023imt}.
This is because ISAC co-designs hardware chains, radio waveforms, signal processing modules, and transmission protocols to enable simultaneous sensing and communication over shared resources effectively.
Consequently, ISAC has demonstrated its potential to efficiently harmonize these dual functionalities and has thus attracted significant global interest~\cite{zhang2021overview,wei2023integrated,liu2022survey,liu2023energy,mu2023noma}.

Despite its advantages, the dual-service provisioning capability of ISAC is becoming increasingly inadequate to meet the concurrent and heterogeneous service demands of emerging scenarios under constrained resources, driven by the rapid integration of wireless communications into diverse vertical applications in future networks.
Therefore,  it is imperative to move beyond the dual-service paradigm of ISAC toward integrated heterogeneous service provisioning (IHSP) \cite{jia2024integrated}, which generalizes ISAC by supporting a broader portfolio of services, including sensing, communication, positioning, computation, control, and other emerging functionalities, within a unified infrastructure.
This holistic paradigm enhances efficiency in energy, spectrum, and hardware utilization, while enabling flexible and scalable cross-service coordination.

ISAC, which is advancing rapidly, can be regarded as a subfield of IHSP.
A critical recent research area of ISAC is the development of resource-sharing schemes to achieve various trade-offs between the key performance indicators (KPIs) of sensing and communication services.
Specifically, the KPIs of sensing services include Cram\'{e}r-Rao bound (CRB)  \cite{liu2021cramer,wei2023carrier,Li2025integeated}, successful detection probability \cite{dong2022sensing},   mutual information rate (MIR) \cite{ni2021multi}, and beampattern mismatch error \cite{hua2023optimal}.
For communication services, the KPIs include the communication rate \cite{hua2023mimo,hu2025drift}, multi-user interference \cite{yu2022precoding}, outage probability \cite{bazzi2023outage}, and bit error rate (BER) \cite{wu2024low}.
In addition, \cite{wu2024low} employed singular value decomposition (SVD) and the Lagrangian dual method to optimize precoding in the delay-Doppler domain, aiming to minimize communication BER under sensing CRB constraints. Subsequently, \cite{mao2024communication} utilized block coordinate descent and the Lagrangian dual transform to optimize beamforming for maximizing the communication-sensing service region.
Furthermore,  \cite{wei2023waveform} employed an information-theoretic approach to design signal waveforms that optimize the weighted sum of communication and sensing MIR.
Moreover, channel temporal correlation has been leveraged in ISAC systems to balance communication rate and sensing CRB \cite{chen2023impact, chen2024learning, liu2020radar, liu2022learning, zhang2024predictive}. For instance, \cite{chen2023impact, chen2024learning} analyzed the impact of channel aging on system performance and optimized channel estimation intervals to reduce training overhead.
Additionally, \cite{liu2020radar} utilized extended Kalman filtering (EKF), while \cite{liu2022learning, zhang2024predictive} adopted deep learning-based methods for predictive beamforming.

However, most existing designs of communication and ISAC systems are inadequate for IHSP and cannot be directly extended to support it due to the following two major issues:
\begin{itemize}
\item Ineffective performance evaluation metrics: conventional designs rigidly enforce diverse KPIs for resource allocation, frequently causing outages for IHSP as they fail to compromise among competing KPI requirements from concurrent users under constrained radio resources.
\item Inefficient multiple access (MA) schemes: conventional designs rely on rigid single-domain or fixed MA schemes \cite{li2025improving,hu2025advanced}, leading to reduced resource efficiency and limited user capacity under dynamic resource situations.
 \end{itemize}

A promising way to address the performance evaluation challenges in IHSP systems is to develop a flexible and comprehensive performance metric.
Therefore, our recent studies \cite{chen2021maximization, li2023value,li2024maximizing} have developed the Value-of-Service (VoS) metric, a novel soft performance evaluation metric designed to capture the value or significance of individual service provisioning events.
By incorporating user-specific KPI demands (e.g., latency, reliability, and transmission rate), the VoS metric enables tailored performance optimization for both end-users and infrastructure providers, offering a compromised and adaptive evaluation of service performance.
Specifically, in \cite{chen2021maximization}, the concept of VoS was introduced to capture the impact of task completion latency and device energy consumption in offloading decisions within mobile collaborative computing networks.
This framework was further developed in \cite{li2023value}, where a specific mathematical expression of VoS was applied in ISAC networks to evaluate sensing accuracy and communication rates for both real-time and delay-tolerant applications.
Subsequently, \cite{li2024maximizing} expanded this approach to guide fairness resource allocation in multi-user collaborative ISAC networks.
However, the existing VoS metrics and conventional designs are developed by considering only one KPI for a single specific service type, neglecting the necessity for IHSP to accommodate multiple service types, as well as a single service with diverse KPIs. Furthermore, they fail to account for user-specific tolerances and the necessary compromises in performance degradation across different KPIs.

Moreover, the recently developed multi-dimensional MA (MDMA) in \cite{liu2020multi,mei2022multi, chen2024otfs} presents a promising solution to address the challenges of inefficient MA.
 Explicitly, MDMA is a hybrid MA technology that flexibly manages interference across various radio resource domains by effectively integrating both orthogonal and non-orthogonal MA schemes opportunistically. It leverages user-specific interference tolerances and cancellation capabilities to allocate the most suitable access method for each user across resource domains.
Specifically, in conventional multi-user multi-input single-output (MISO) communication systems,  \cite{liu2020multi} leveraged MDMA to adaptively multiplex coexisting devices across frequency, time, space, power, and code domains, thereby maximizing service performance while minimizing non-orthogonality between users.
This approach was extended in \cite{mei2022multi} to incorporate resource utilization costs, considering interference introduced and device capability, for individualized service provisioning.
Moreover, \cite{chen2024otfs} utilized MDMA as a foundational platform to flexibly multiplex coexisting users across power, space, and delay-Doppler domains in an orthogonal time-frequency space (OTFS) system.
However, these MDMA schemes, which were designed based on the KPI of communication rate for communication systems,  are not suitable for IHSP systems due to diverse KPIs and the complex functional properties involved in analysis and optimization.

Accordingly, we address these challenges and advance toward IHSP by integrating two elastic design strategies: compromised service value assessment and flexible multi-dimensional resource multiplexing. 
Then, we propose a VoS-prioritized elastic MDMA mechanism for a multi-user IHSP system.  The main contributions are summarized as follows:

\begin{itemize}
\item
We modify the existing VoS metric into a more general formulation that incorporates both range and slope elastic parameters. This enhanced metric captures user-specific elasticity in tolerating performance degradation under constrained resources.
    Here, the range elasticity parameters regulate the meaningful range of KPIs, while the slope elasticity parameter determines the impact of performance loss compromises on the value of service provisioning. This enhanced VoS serves as a foundation for prioritizing and enabling effective service provisioning among competing services.

\item
Based on the modified VoS metric, we adapt the MDMA to elastically multiplex services across time, frequency, space, and power domains using appropriate MA schemes.
This protocol leverages user-specific interference tolerances and cancellation capabilities across different resource domains to reduce resource conflicts and co-channel interference within the same domain.

\item
We maximize the proportional VoS of all users by jointly optimizing the MDMA for assigning services to resource bins, along with power allocation.
However, this is a mixed-integer nonlinear programming (MINLP) problem and is challenging to solve. Therefore,  we first propose a monotonic optimization-aided dynamic programming (MODP) algorithm to find its optimal solution.
Then, we develop a VoS-prioritized successive convex approximation (SCA) sub-optimal algorithm that uses VoS prioritization for MDMA design and SCA for power allocation.

\end{itemize}

Organizations:
Section~\ref{Sec2} introduces the IHSP system model.
Section~\ref{SecDerivation} derives the performance metrics of the system.
Section~\ref{Sec4}  formulates the optimization problem and transforms it into an equivalent tractable formulation to facilitate algorithm development.
Section~\ref{Sec5} and  Section~\ref{Sec6} propose optimal and suboptimal algorithms, respectively, to efficiently solve the reformulated problem.
Finally, Section~\ref{Sec7}  provides  simulation results and Section~\ref{Sec8}  concludes the paper.

Notation: Scalars, vectors, and matrices are represented by lowercase, bold lowercase, and bold uppercase letters, respectively, i.e., $a$, $\mathbf{a}$, and $\mathbf{A}$. Sets and their cardinalities are denoted using blackboard bold font and vertical bars, i.e., $\mathbb{K}$ and $|\mathbb{K}|$, respectively. The transpose and conjugate transpose operations are denoted as $(\cdot)^{\rm T}$ and $(\cdot)^{\rm H}$, respectively, while ${\rm E}(\cdot)$ represents the expectation operator. Finally,  ${\mathbb{R}}_{+}^{D}$ represents the set of $D$-dimensional positive real numbers,   ${\mathbb{C}}$ represents the set of complex numbers; and  ${\cal CN}(\mu,\sigma)$ represents the circularly symmetric complex Gaussian (CSCG) distribution with mean $\mu$ and covariance  $\sigma$.

\section{System Model}\label{Sec2}

\begin{figure}[t]
\centering
   \includegraphics[width=.43\textwidth]{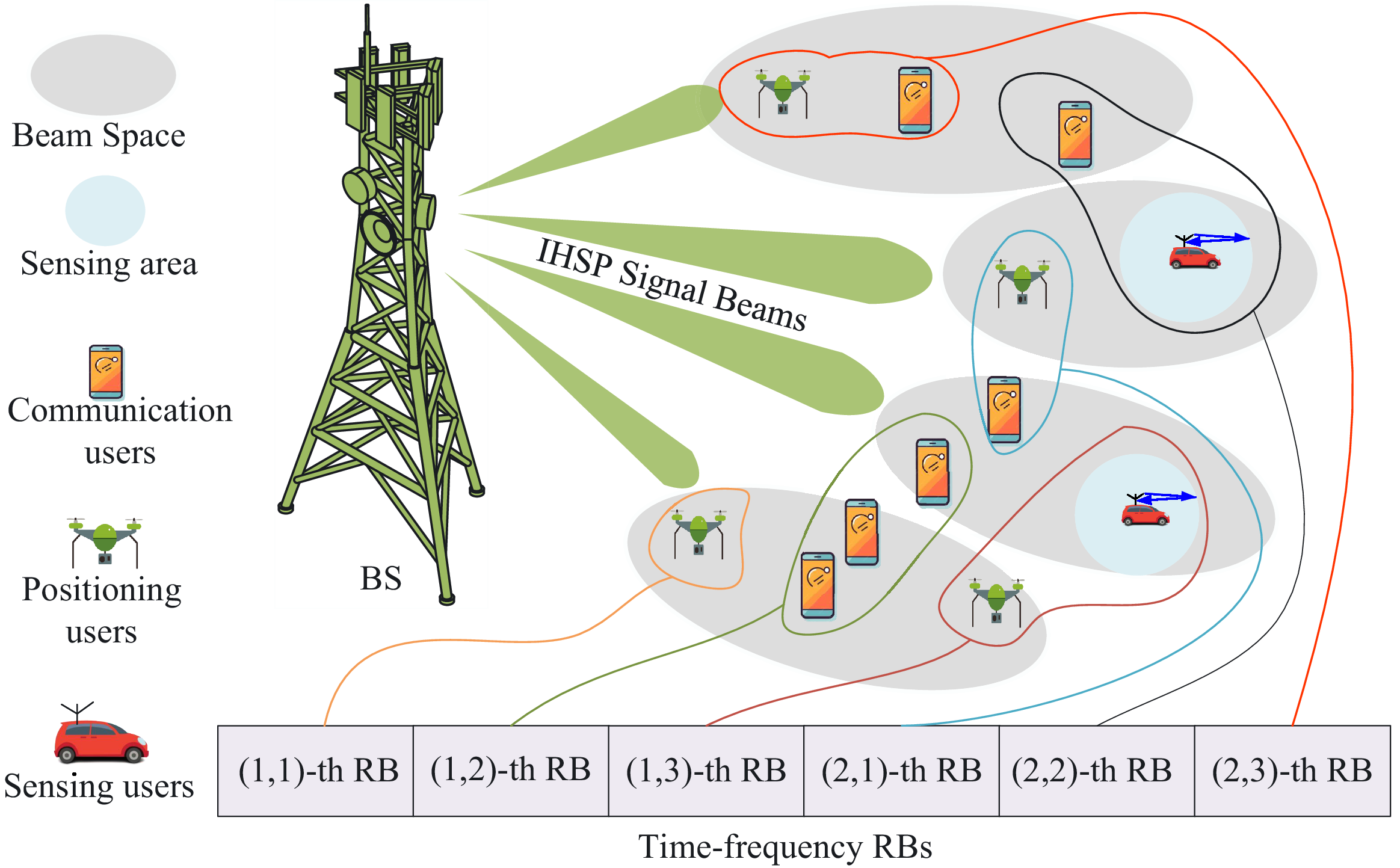}
  \caption{An illustration of MDMA for IHSP system, where  $(m, n)$-th RB refers to the RB located on the $m$-th frequency sub-band and the $n$-th time sub-frame, with $M=2$ and $N=3$.}\label{figSys1}
 \end{figure}

As shown in Fig. \ref{figSys1}, we consider a multi-user IHSP system consisting of one full-duplex BS equipped with $L_{\rm tx}$ antennas and $K$ single-antenna users with index set ${\mathbb K} = \left\{ {1,2,\ldots, K} \right\}$.
 Each user is associated with a specific service type and classified into three groups:
communication users, who receive the downlink independent information from the BS; positioning  users, who reflect the positioning signals
from the BS to enable round-trip positioning; and sensing users, who request the BS to schedule resource blocks (RBs) for transmitting their own predetermined sensing signals to detect the presence of a potential nearby target.
Specifically, we use Type-${\rm{X}}$, where ${\rm X} \in \left\{ {\rm{C}}, {\rm{P}}, {\rm{S}} \right\}$, to represent the service type: ${\rm X}={\rm C}$ for communication services, ${\rm X}={\rm P}$ for positioning services, and ${\rm X}={\rm S}$ for sensing services.
Moreover, we denote the index set of users requiring Type-$\rm X$ service  by ${{\mathbb K}_{\rm{X}}}$, and we assume
${{\mathbb K}_{\rm{C}}} = \left\{ { {1, \ldots ,\left| {{{\mathbb K}_{\rm C}}} \right|}} \right\}$, ${\mathbb K}_{\rm P} =\left\{ { {\left| {{{\mathbb K}_{\rm C}}} \right| + 1, \ldots ,\left| {{{\mathbb K}_{\rm C}}} \right| + \left| {{{\mathbb K}_{\rm  P}}} \right|} } \right\}$, and ${{\mathbb K}_{\rm{S}}} = \left\{ { {\left| {{{\mathbb K}_{\rm C}}} \right| + \left| {{{\mathbb K}_{\rm P}}} \right| + 1, \ldots ,K} } \right\}$, respectively.

\subsection{Elastic VoS Definition}

In this paper, we assume that users requiring the same service type have different KPI requirements within the same KPI set, whereas users requesting different service types have distinct KPI demands across different KPI sets.
For example, Type-$\rm C$  users may exhibit varying requirements for the KPI set, including communication rate and latency,
whereas Type-$\rm P$ users may have different requirements for another KPI set, including range, angle, and velocity estimation accuracies.

Next,  we assume that each user has a specific tolerance level for performance degradation between the desired and achieved KPI values under constrained resources.
Consequently, we develop an elastic VoS metric to evaluate the value of completing a service while considering user-specific   performance degradation  tolerances
across specific KPIs.
To achieve this, we define a general value normalization function  ${\cal V}(\cdot)$ for any type of KPI, which incorporates user-specific elasticity with respect to performance loss on that KPI.
Mathematically,  the normalized value for the $i$-th KPI of user $k$ demanding Type-$\rm X$ service, considering user-specific elasticity, is defined as:
\begin{align}
&{V}_{k}^{{\rm X},i} \buildrel \Delta \over = {\cal V}\left( {Q_{k}^{{\rm X}, i},\tilde Q_{k}^{{\rm X}, i},\alpha _{k}^{{\rm X}, i},\beta _{k}^{{\rm X},i}} \right),
\nonumber \\
&\qquad\qquad{\rm for}\;1\le i\le  \left| {{\mathbb Q}^{\rm{X}}} \right|, k \in {{\mathbb K}_{\rm{X}}}\;    \& \; {\rm{X}} \in \left\{ {{\rm{C}},{\rm{P}},{\rm{S}}} \right\}, \label{eqQos2}
\end{align}
where ${{\mathbb Q}^{\rm{X}}}$ represents the set of KPIs required in Type-$\rm X$ service, and $\left| {{\mathbb Q}^{\rm{X}}} \right|$ denotes the number of KPI types in this set. Besides, ${{ Q}}_{k}^{{\rm X},i}$ and ${\tilde{ Q}}_{k}^{{\rm X},i}$ represent the achieved  and desired values of the $i$-th KPI of Type-$\rm X$ service at user $k$, respectively.
Moreover, $\alpha_{k}^{{\rm X}, i}$ and $\beta_{k}^{{\rm X}, i}$ represent the user-specific elasticity parameters, which characterize the elasticity of performance loss between ${\tilde{Q}}_{k}^{{\rm X}, i}$ and $Q_{k}^{{\rm X}, i}$, and control the impact of this performance loss on the normalized value.

\begin{figure}[t]
   \centering
   \includegraphics[width=.43\textwidth]{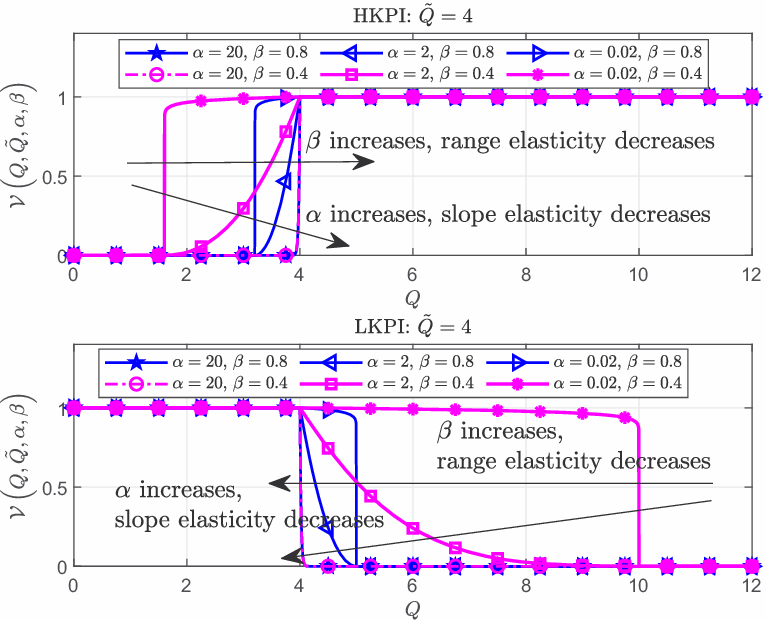}
  \caption{An illustration of the value normalization function.}\label{figSys2}
 \end{figure}

In general, although there are many types of KPIs, they can be classified into two categories: High-KPI (HKPI), where higher values indicate better performance (e.g., rate and detection probability); Low-KPI (LKPI), where lower values are preferable (e.g., latency and CRB). Upon assuming $\alpha>0$ and $0<\beta< 1$, as illustrated in Fig. \ref{figSys2}, function ${\cal V}\left( {Q,\tilde Q,\alpha ,\beta } \right)$  can be defined as follows:
\begin{itemize}
\item For HKPI, we have
\begin{align} \label{equationQos2a}
\!\!\!&{\cal V}\left( {Q,\tilde Q,\alpha ,\beta } \right) \nonumber\\
\!\!\!=&\left\{ {\begin{array}{*{20}{l}}
{1,{\rm{if}}\;\tilde Q < Q,}\\
\!\!{{{\!\!\left( \!{\frac{1}{{{A_{{\rm H}}}}}\!\left( {\frac{1}{{1 + {e^{ - \alpha \left( {\frac{Q}{{\tilde Q}} - 1} \right)}}}} - {B_{{\rm H}}}} \right)}\!\! \right)}^\alpha }\!\!  {\rm{,  if}}\;\beta \tilde Q\! \le\! Q\! \le\! \tilde Q,}\\
{0,{\rm{if}}\;Q < \beta \tilde Q,}
\end{array}} \right.
\end{align}
where ${B_{\rm H}} = \frac{1}{{1 + {e^{ - \alpha \left( {\beta  - 1} \right)}}}}$ and ${A_{\rm H}} = \frac{1}{2} - {B_{{\rm H}}}$.
 \item For LKPI, we have
 \begin{align}\label{equationQos2b}
\!\!\!&{\cal V}\left( {Q,\tilde Q,\alpha ,\beta } \right)\nonumber\\
\!\!\!=&\left\{ {\begin{array}{*{20}{l}}
{0,{\rm{if}}\;\frac{{\tilde Q}}{\beta } < Q,}\\
\!\!{{{\!\!\left( {\frac{1}{{{A_{\rm L}}}}\left( {\frac{1}{{1 + {e^{\alpha \left( {\frac{Q}{{\tilde Q}} - 1} \right)}}}} - {B_{\rm L}}} \right)}\!\! \right)}^\alpha }\!\!{\rm{, if}}\;\tilde Q \!\le \!Q \!\le\! \frac{{\tilde Q}}{\beta },}\\
{1,{\rm{if}}\;Q < \tilde Q,}
\end{array}} \right.
\end{align}
where ${B_{\rm L}} = \frac{1}{{1 + {e^{\alpha \left( {\frac{1}{\beta } - 1} \right)}}}}$ and ${A_{\rm L}} = \frac{1}{2} - {B_{\rm L}}$.

\end{itemize}

In \eqref{equationQos2a} and \eqref{equationQos2b}, $B_{\rm H}$ and $B_{\rm L}$ are the constant shift parameters, and $A_{\rm H}$ and $A_{\rm L}$ are the constant normalization parameters.
As shown in  Fig. \ref{figSys2},  $\beta$ can be called the range elasticity parameter, which constrains the value to zero when  the achieved  KPI is smaller than the worst threshold for HKPI, i.e., $Q < \beta \tilde{Q}$, or when the achieved  KPI is larger than the worst threshold for LKPI, i.e.,
 $Q \ge \frac{\tilde{Q}}{\beta}$. Here, a larger $\beta$  indicates reduced elasticity in the meaningful range of the achieved  KPI on the normalized value.
Besides, $\alpha$ can be called the slope elasticity parameter, which controls the slope of the effect that the performance loss of a given KPI has on the final normalized value.
Additionally, as $\alpha$ increases from zero to infinity, the function transforms from a log-concave shape to a sigmoid shape. Thus, a larger $\alpha$ results in a steeper slope, leading to greater performance degradation in terms of the normalized value.

Finally, the elastic VoS for Type-$\rm X$ service at user $k$ is evaluated as a weighted-proportional function of all ${V}_{k}^{{\rm X},i}$, i.e.,
\begin{align}
V_k^{\rm{X}}  \buildrel \Delta \over =\prod\nolimits_{i = 1}^{\left| {{{\mathbb Q}^{\rm{X}}}} \right|} {{{\left( {V_k^{{\rm{X}},i}} \right)}^{w_k^{{\rm{X}},i}}}},
 \label{eqVos}
\end{align}
where $w_k^{{\rm{X}},i}$ is the constant fairness weight of  ${V}_k^{{\rm{X}},i}$ defined in~\eqref{eqQos2}.
Note that the defined elastic VoS is a unified metric that is also applicable to other service types, such as control or synchronization services \cite{jia2019distributed}.

\subsection{ Elastic MDMA Design}

To support more services and enhance the VoS of IHSP systems, we propose elastic MDMA, which leverages multi-dimensional resource diversity across time, frequency, power, and spatial domains, as well as service-specific interference cancellation capabilities, to increase the number of multiplexed services within each orthogonal RB.
Specifically, the time-frequency (TF) resources are divided into $M \times N$ RBs, where $M$ and $N$ represent the numbers of sub-bands and time sub-frames, respectively.
As shown in Fig. \ref{figSys1},  each TF RB can accommodate one service as orthogonal MA (OMA) or multiple services as non-orthogonal MA (NOMA) based on the channel quality or interference level with respect to other users in power domains.
Specifically, the $(m, n)$-th RB refers to the RB located on the $m$-th frequency sub-band and the $n$-th time sub-frame.
Then,  the MDMA design is denoted by binary variables, i.e.,  ${ a}_{kmn}\in\left\{0,1\right\}$, where ${ a}_{kmn}=1$ if the service of user $k$ is accommodated in the $(m,n)$-th RB, and ${ a}_{kmn}=0$ otherwise. Moreover, each user service is assumed to be assigned to only one RB.
Note that OMA inherently avoids interference, whereas NOMA may introduce interference, which can be neglected, considered, or mitigated via self-interference cancellation (SIC), depending on the specific multiplexed service types and channel conditions \cite{mei2022multi}.
By implementing MDMA, resource conflicts and co-channel interference are effectively minimized, thereby supporting diverse requirements in IHSP systems.

\section{Elastic VoS Derivations in IHSP}\label{SecDerivation}
 This section characterizes the detailed formula of VoS for Type-$\rm X$ service, which will be used to guide the elastic MDMA design in IHSP.

\subsection{VoS  of Communication Service}

For \mbox{Type-$\rm C$} service, we focus on two KPIs:  the transmission rate (HKPI) and the service completion latency (LKPI), i.e., ${{\mathbb Q}^{\rm{C}}} = \left\{ {{\rm{transmission}}\;{\rm{rate}},\;{\rm{service}}\;{\rm{latency}}} \right\}$.

Firstly, let ${\bf{h}}_{kmn}\in{\mathbb C}^{L_{\rm tx}\times 1}$ represent the downlink channel response between the BS and user $k$ in the $(m,n)$-th RB, and let ${g}_{kk'mn}$ represent the channel response between users $k$ and $k'$ in the same RB.
Next,
we assume that each RB consists of $B$ subcarriers and $L$ symbols, where the bandwidth of each subcarrier is  $\Delta_f$ and the symbol duration is $T$.
Then, let $s_{kmn}^{bl}\sim{\cal CN}\left(0,1\right)$  represent the orthogonal frequency division multiplexing (OFDM) symbol on subcarrier $b$ and time slot $l$  within the $(m,n)$-th RB for service $k$.
Note that $s_{kmn}^{bl}$  is transmitted by the BS for user ${k \in {{\mathbb K}_{\rm{C}}} \cup {{\mathbb K}_{\rm{P}}}}$  when requesting  Type-$\rm C$ and Type-$\rm P$ services, or transmitted by user $k\in {{\mathbb K}_{\rm{S}}} $ when demanding \mbox{Type-$\rm S$} services.
Consequently, if $a_{kmn}=1$ for $k\in {\mathbb K}_{\rm C}$, the received ($b,l$)-th signal at user $k$ within the ($m,n$)-th RB can be generally expressed as:
\begin{align}
y_{kmn}^{{\rm{C}},bl}& = \underbrace {\sqrt {{p_{kmn}}} {\bf{h}}_{kmn}^{\rm{H}}{\bf{w}}_{kmn}^{{\rm{tx}}}s_{kmn}^{bl}}_{{\rm{Desired}}\;{\rm{signal}}}  \nonumber\\
 &+ \underbrace {{\bf{h}}_{kmn}^{\rm{H}}\sum\limits_{k' \ne k,k' \in {{\mathbb K}_{\rm{C}}} \cup {{\mathbb K}_{\rm{P}}}} {{a_{k'mn}}\sqrt {{p_{k'mn}}} {\bf{w}}_{k'mn}^{{\rm{tx}}}s_{k'mn}^{bl}} }_{{\rm{Interference}}\;{\rm{from}}\;{\rm{the}}\;{\rm{BS}}\;}\nonumber\\
 &+ \underbrace {\sum\nolimits_{k' \in {{\mathbb K}_{\rm{S}}}} {{a_{k'mn}}\sqrt {{p_{k'mn}}} {g_{kk'mn}}s_{k'mn}^{bl}} }_{\;{\rm{Interference}}\;{\rm{from}}\;{\rm{sensing}}\;{\rm{users}}} + u_{kmn}^{{\rm{C}},bl},
 \end{align}
where ${\bf{w}}_{kmn}^{\rm tx}\in{\mathbb C}^{L_{\rm tx} \times 1}$ and ${p_{kmn}}$ for $k \in {{\mathbb K}_{\rm{C}}} \cup {{\mathbb K}_{\rm{P}}}$ are the transmit beamforming vectors and allocated powers at the BS of the communication/positioning signal $s_{kmn}^{bl}$, respectively.  Besides, ${p_{kmn}}$ for $k \in {{\mathbb K}_{\rm{S}}}$ is the transmission power at user $k$ for its own sensing signal $s_{kmn}^{bl}$.
Moreover, $u_{kmn}^{{\rm{C}},bl} \sim {\cal CN}\left(0,\sigma_k\right)$  is the corresponding received  noise term at user $k$ with mean zero and covariance $\sigma_k$.

Note that the interference from positioning and sensing signals $s_{kmn}^{bl}$ for ${k \in {{\mathbb K}_{\rm{S}}} \cup {{\mathbb K}_{\rm{P}}}}$ can be cancelled because these signals are prior known sequences at the BS and communication users.
As for the interference cancellation from other communication signals $s_{k'mn}^{bl}$ where $ k' \ne k$ and $k'\in {{\mathbb K}_{\rm{C}}}$,
we can apply the SIC utilized in the conventional power-domain NOMA.
Specifically, the near user performs SIC to remove the interference from far users before decoding its own signal, whereas the far user treats the signals from near users as interference and directly decodes its own signal. Specifically, we assume that the distances between the BS and communication users increase with their indices, resulting in a decrease in channel powers as the user indices increase, i.e.,
\begin{align}
{\left\| {{{\bf{h}}_{k'mn}}} \right\|^2} \le {\left\| {{{\bf{h}}_{kmn}}} \right\|^2}, \;{\rm for}\; 1 \le k <  k' \le \left| {{{\mathbb K}_{\rm C}}} \right|,\forall m,n.
 \end{align}
Based on the SIC principle,  we must ensure that the signal intended for user $k$ can be successfully decoded by all its relative near users sharing the same RB. This implies that the signal-to-interference-plus-noise-ratio (SINR)  of the signal intended for user $k$ satisfies the following condition \cite{al2019energy}:
 \begin{align}
z_{kmn}^{\rm{C}} = \min \left\{ {\gamma _{qk}^{mn}\left| {{a_{qmn}} = 1,1 \le q \le k \le \left| {{{\mathbb K}_{\rm{C}}}} \right|} \right.} \right\},
 \end{align}
where $\gamma _{qk}^{mn}$ is the SINR of decoding the signal intended for user $k$ at user $q$, i.e.,
\begin{align}
\gamma _{qk}^{mn} = \frac{{{p_{kmn}}\chi _{qkmn}^{\rm{C}}}}{{\sum\nolimits_{j = 1}^{k - 1} {{a_{jmn}}{p_{jmn}}\chi _{qjmn}^{\rm{C}}}  + {\sigma _q}}}.
 \end{align}
Here, we define  $\chi _{qkmn}^{\rm{C}} = {\left| {{\bf{h}}_{qmn}^{\rm{H}}{\bf{w}}_{kmn}^{{\rm{tx}}}} \right|^2}$ for $q,k\in{\mathbb K}_{\rm C}$.

Next, the achieved  performances of the first and second  KPIs of \mbox{Type-$\rm C$} service, i.e., transmission rate  and  latency of service $k$, on the $(m,n)$-th RB can be expressed as:
\begin{align}
Q_{kmn}^{{\rm{C}},1} \buildrel \Delta \over = \log_2 \left(1 + z _{kmn}^{\rm C} \right),\;Q_{kmn}^{{\rm{C}},2} \buildrel \Delta \over = nLT. \label{eqQC}
 \end{align}

From \eqref{eqVos}, the VoS of user $k \in {\mathbb K}_{\rm C}$ is expressed as 
\begin{equation}
V_k^{\rm{C}} = \sum\nolimits_{m = 1}^M {\sum\nolimits_{n = 1}^N {{a_{kmn}}\prod\nolimits_{i = 1}^2 {{{\left( {V_{kmn}^{{\rm{C}},i}} \right)}^{w_{k}^{{\rm{C}},i}}}} } },\label{eqVKC}
\end{equation}
where $V_{kmn}^{{\rm{C}},i} = {\cal V}\left( {Q_{kmn}^{{\rm{C}},i},\tilde Q_{k}^{{\rm{C}},i},\alpha _k^{{\rm{C}},i},\beta _k^{{\rm{C}},i}} \right)$ is calculated by
\eqref{eqQos2}-\eqref{equationQos2b} and ${w_{k}^{{\rm{C}},i}}$ is defined after \eqref{eqVos}.

\subsection{VoS of Positioning Service}
Let  ${\bf{x}}_k = \left[ {\theta_k,d_k,v_k} \right]^{\rm T}$ represent the parameters of interest in the positioning service for $k \in {\mathbb K}_{\rm P}$, where $\theta_k$, $d_k$, and $v_k$ denote the corresponding angle, distance, and velocity relative to the BS, respectively.
Then, for Type-$\rm P$ service, we consider four LKPIs: the CRBs of angle, distance, and velocity, and the service completion latency, i.e., ${{\mathbb Q}^{\rm{P}}} = \left\{ {{\rm{angle/distance/velocity}}\;{\rm{CRBs}},\;{\rm{service}}\;{\rm{latency}}} \right\}$.

Then, we assume that the self-interference at the full-duplex BS is perfectly eliminated \cite{xiao2022waveform,li2024full}\footnote{We note that while the residual interference from imperfect cancellation may marginally degrade performance, it does not impede the algorithm's implementation. Designing robust resource allocation for such non-ideal scenarios is a promising avenue for future research.}. Besides, we focus solely on the echo signals reflected by positioning users, as they comparatively have larger radar cross-sections (RCSs), while ignoring the echo signals from communication and sensing users due to their comparatively smaller RCSs and larger distances.
Then,  if $a_{kmn}=1$ for $k\in {\mathbb K}_{\rm P}$, the received ($b,l$)-th signal at the BS within the ($m,n$)-th RB can be generally  given by:
\begin{align}
{\bf{y}}_{mn}^{{\rm{P}},bl} &= \sum\limits_{k' \in {{\mathbb K}_{\rm{P}}}} {{\rho_{k'mn}}e^{{\rm j}{{\rm \varphi}_{k'mn}}}{\bf{v}}\left( {{\theta _{k'}}} \right){e^{{\rm{j}}2\pi \left( {{\nu_{k'm}}lT - b{\Delta _f}{\tau _{k'}}} \right)}}\bar s_{k'mn}^{bl}} \nonumber \\
 &+ \underbrace {\sum\limits_{k' \in {{\mathbb K}_{\rm{S}}}} {{a_{k'mn}}\sqrt{{p_{k'mn}}} {\bf{h}}_{k'mn}s_{k'mn}^{bl}} }_{ {\rm{Interference}}\;{\rm{from}}\;{\rm{sensing}}\;{\rm{users}}} + {\bf{u}}_{mn}^{{\rm{P}},bl},\label{eq9}
\end{align}
where the signal $\bar s_{k'mn}^{bl}$ denotes the effective reflected signal by target $k'$, i.e.,
\begin{align}
\bar s_{k'mn}^{bl} \!=\! {\bf{v}}\left( {{\theta _{k'}}} \right)^{\rm{H}} \!{\sum\nolimits_{i \in {{\mathbb K}_{\rm{C}}} \cup {{\mathbb K}_{\rm{P}}}} {{a_{imn}}\sqrt {{p_{imn}}} {\bf{w}}_{imn}^{{\rm{tx}}}s_{imn}^{bl}} },
\end{align}
with ${\rm{E}}\left( {{{\left| {\bar s_{k'mn}^{bl}} \right|}^2}} \right) = \sum\nolimits_{i\in {{\mathbb K}_{\rm{C}}} \cup {{\mathbb K}_{\rm{P}}}} {{a_{imn}}{p_{imn}}\chi _{k'imn}^{\rm{P}}} $ and $\chi _{k'imn}^{\rm{P}} = {\left| {{\bf{v}}{{\left( {{\theta _{k'}}} \right)}^{\rm{H}}}{\bf{w}}_{imn}^{{\rm{tx}}}} \right|^2}$ for $k'\in{\mathbb K}_{\rm P}$ and $i\in{\mathbb K}_{\rm C}\cup{\mathbb K}_{\rm P}$.
Besides,  ${\bf{v}}\left( {{\theta }} \right)={\left[ {1,{e^{j\pi \sin \theta }},\cdots,{e^{j\pi \left( {{L_{\rm tx}} - 1} \right)\sin \theta }}} \right]^{\rm H}} $ is the steering
vector with angle $\theta$ when assuming half-wavelength antenna spacing.
Moreover, ${\rho _{kmn}} = \sqrt {\frac{{c_o^2\delta _k^{{\rm{RCS}}}}}{{{{\left( {4\pi } \right)}^3}f_m^2d_k^4}}} $, ${{\rm \varphi}_{kmn}}$, ${\tau _k} = \frac{{2{d_k}}}{c_o}$, and ${\nu _{km}} = \frac{{2{v_k}{f_m}}}{c_o}$, respectively, represent the round-trip attenuation factor, phase noise, time delay,  and Doppler phase shift with respect to user $k$ for $k\in{\mathbb K}_{\rm P}$ on the $(m,n)$-th RB; $T$ and $c_o$, respectively, represent the OFDM symbol duration including the cyclic prefix and the speed of light;  ${\bf{u}}_{mn}^{{\rm{P}},bl}$ is the received noise.
Besides, $\delta _k^{{\rm{RCS}}}$  is the corresponding RCS and  ${f _{m}}=mB\Delta_f+f_c$, where ${f_c}$ is  carrier frequency.

Due to the prior known sensing signals, the corresponding interference can be cancelled. Then,  the signal in \eqref{eq9} received by the $\ell$-th antenna at the BS is rewritten as \begin{align}
\bar y_{mn}^{{\rm{P}},bl\ell }= & \sum\limits_{k' \in {{\mathbb K}_{\rm{P}}}} {{\rho _{k'mn}}{e^{{\rm j}{\varphi _{k'mn}}}}{e^{{\rm{j}}\pi \left[ {2\left( {{\nu _{k'}}lT - b{\Delta _f}{\tau _{k'}}} \right) - \ell \sin {\theta _{k'}}} \right]}}\bar s_{k'mn}^{bl}} \nonumber\\
 &+ u_{mn}^{{\rm{P}},bl\ell }, \label{eqCRLBsignalA}
\end{align}
where $u_{mn}^{{\rm{P}},bl\ell} \sim {\cal CN}\left(0,\sigma_0\right)$ is the equivalent receiving noise at the BS  with mean zero and covariance $\sigma_0$.

\begin{theorem}\label{theorem1}
Assuming the positioning users are well-separated in the plane, upon denoting $z _{kmn}^{\rm{P}} = \frac{{\sum\nolimits_{k' \in {{\mathbb K}_{\rm{C}}} \cup {{\mathbb K}_{\rm{P}}}} {{a_{k'mn}}{p_{k'mn}}\chi _{kk'mn}^{\rm{P}}} }}{{{\sigma _0}}}$,  the CRBs of angle $\theta_k$ (i.e., $Q_{kmn}^{{\rm P},1}$), distance $d_k$ (i.e., $Q_{kmn}^{{\rm P},2}$), and velocity $v_k$ (i.e., $Q_{kmn}^{{\rm P},3}$) can be generally approximated by:
\begin{subequations}\label{eq12}
\begin{align}
{\rm{E}}\left( {{{\left| {{\theta _k} - {{\hat \theta }_k}} \right|}^2}} \right) &\ge \frac{{I_{kmn}^{{\rm{P}},\theta }}}{{z _{kmn}^{\rm P}}} = Q_{kmn}^{{\rm P},1},\\
{\rm{E}}\left( {{{\left| {{d_k} - {{\hat d}_k}} \right|}^2}} \right) &\ge \frac{{I_{kmn}^{{\rm{P}},d}}}{{z _{kmn}^{\rm P}}}= Q_{kmn}^{{\rm P},2} ,\\
{\rm{E}}\left( {{{\left| {{v_k} - {{\hat v}_k}} \right|}^2}} \right) &\ge \frac{{I_{kmn}^{{\rm{P}},v}}}{{z _{kmn}^{\rm P}}}= Q_{kmn}^{{\rm P},3} ,
\end{align}
\end{subequations}
where ${I_{kmn}^{{\rm{P}},\theta } = \frac{1}{2}\left[ {{{\bf{J}}_{kmn}}} \right]_{11}^{ - 1}}$, ${I_{kmn}^{{\rm{P}},d} = \frac{{c_o^2}}{{32{{\left( {\pi {\Delta _f}} \right)}^2}}}\left[ {{{\bf{J}}_{kmn}}} \right]_{22}^{ - 1}}$, and ${I_{kmn}^{{\rm{P}},v} = \frac{{c_o^2}}{{32{{\left( {\pi T{f_m}} \right)}^2}}}\left[ {{{\bf{J}}_{kmn}}} \right]_{33}^{ - 1}}$.  Here, $\left[ {{{\bf{J}}_{kmn}}} \right]_{ii}^{ - 1}$ for $1\le i\le 3$ is the $(i,i)$-th element of matrix ${{{\left[ {{{\bf{J}}_{kmn}}} \right]}^{ - 1}}}$, where
\begin{align}
{{\bf{J}}_{kmn}} &=  \sum\nolimits_{\ell  = 0}^{{L_{{\rm{tx}}}} - 1} {\sum\nolimits_{b = 0}^{B - 1} {\sum\nolimits_{l = 0}^{L - 1} {{\bf{J}}_{kmn}^{bl\ell }} } }\in{\mathbb C}^{3\times3},\\
{\bf{J}}_{kmn}^{bl\ell }& =\rho _{kmn}^2\times \nonumber\\
&\left[ {\begin{array}{*{10}{c}}
\!\!\!{{{\left({\ell \cos {\theta _k}} \right)}^2}}&\!\!\!{b\ell \cos {\theta _k}}  &\!\!\!{\ell l\cos {\theta _k}}   &\!\!\!\!\!\!\!0     &\!\! \!\!\!\!\!\!{\ell \cos {\theta _k}}\\
\!\! \!{b\ell \cos {\theta _k}}                    &\!\!\!{{b^2}}                   &\!\!\!{bl}                       &\!\!\!\!\!\!\!0      &\!\!\!\!\!\!\!\!b\\
\!\! \!{\ell l\cos {\theta _k}}                    &\!\!\!{bl}                      &\!\!\!{{l^2}}                    &\!\!\!\!\!\!\!0      &\!\!\!\!\!\!\!\!l\\
\!\! \!0                                           &\!\!\!0                         &\!\!\!0                          &\!\!\!\!\!\!\!{\frac{1}{{\rho _{kmn}^2}}}   &\!\!\!\!\!\!\!\!0\\
\!\! \!{\ell \cos {\theta _k}}                     &\!\!\!b                         &\!\!\!l                           &\!\!\!\!\!\!\!0                  &\!\!\!\!\!\!\!\!\!\!1
\end{array}} \right]. \label{eqJkmn}
\end{align}
\end{theorem}
\begin{IEEEproof} Please refer to Appendix~\ref{appendix1}.
\end{IEEEproof}

Moreover, we  denote the latency of Type-$\rm P$ service for user $k$ within the $(m,n)$-th RB by $Q_{kmn}^{{\rm P},4}  \buildrel \Delta \over =  nLT$.
Then, from \eqref{eqVos},  the VoS of user $k \in {\mathbb K}_{\rm P}$ is expressed as
\begin{equation}
V_k^{\rm{P}} = \sum\nolimits_{m = 1}^M {\sum\nolimits_{n = 1}^N {{a_{kmn}}\prod\nolimits_{i = 1}^4 {{{\left( {V_{kmn}^{{\rm{P}},i}} \right)}^{w_{k}^{{\rm{P}},i}}}} } } ,\label{eqVKP}
\end{equation}
where $V_{kmn}^{{\rm{P}},i} = {\cal V}\left( {Q_{kmn}^{{\rm{P}},i},\tilde Q_{k}^{{\rm{P}},i},\alpha _k^{{\rm{P}},i},\beta _k^{{\rm{P}},i}} \right)$ is calculated by
\eqref{eqQos2}-\eqref{equationQos2b} and ${w_{k}^{{\rm{P}},i}}$ is defined after \eqref{eqVos}.

\subsection{VoS of Sensing Service }

For \mbox{Type-$\rm S$} service, we consider two KPIs: the detection probability (HKPI) and the service completion latency (LKPI), i.e., ${{\mathbb Q}^{\rm{S}}} = \left\{ {{\rm{detection}}\;{\rm{probability}},\;{\rm{service}}\;{\rm{latency}}} \right\}$.

Then, if  $a_{kmn}=1$ for $k\in{\mathbb K}_{\rm S}$,  user $k$ transmits its own sensing signals in the $(m,n)$-th RB and examines the corresponding echoes to detect the presence of a potential nearby target in the desired range-Doppler (RD) bin $\left(  r_k, \nu_k  \right)$.
 Here, $r_k$  and $\nu_k$ when $k\in{\mathbb K}_{\rm S}$ represent the range and Doppler, respectively, between user $k$ and the potential target.
Therefore,  a hypothesis test is performed to detect whether a target exists within  the RD bin $\left( r_k, \nu_k \right)$ for user $k$ ($k\in{\mathbb K}_{\rm S}$):
\begin{itemize}
\item ${{\cal H}_k^0}$(null hypothesis):  no target exists in   bin $\left( r_k ,\nu_k \right)$.
\item ${{\cal H}_k^1}$(alternative hypothesis): a target exists in  bin $\left( r_k ,\nu_k  \right)$.
\end{itemize}

Under the alternative hypothesis, the received ($b,l$)-th echo at user $k\in{\mathbb K}_{\rm S}$ within the ($m,n$)-th RB is generally given by
\begin{align}
y_{kmn}^{{\rm{S}},bl}& = \sqrt {{p_{kmn}}} {\rho _{kmn}}e^{{\rm j}{{\rm \varphi}_{kmn}}}\underbrace {{e^{{\rm{j}}2\pi \left( {{\upsilon _k}lT - b{\Delta _f}\frac{{2{r_k}}}{{{c_0}}}} \right)}}s_{kmn}^{bl}}_{\Omega _{kmn}^{{\rm{1}},bl}}   \nonumber\\
& + \underbrace {{\bf{h}}_{kmn}^{\rm{H}}\sum\limits_{k' \in {{\mathbb K}_{\rm{C}}} \cup {{\mathbb K}_{\rm{P}}}} {{a_{k'mn}}\sqrt {{p_{k'mn}}} {\bf{w}}_{k'mn}^{{\rm{tx}}}s_{k'mn}^{bl}} }_{\Omega _{kmn}^{2,bl}:\;{\rm{Interference}}\;{\rm{from}}\;{\rm{the }}\;{\rm BS}}\nonumber\\
 &+ \underbrace {\sum\limits_{k' \ne k,k' \in {{\mathbb K}_{\rm{S}}}} {{a_{k'mn}}\sqrt {{p_{k'mn}}} {g_{kk'mn}}s_{k'mn}^{bl}} }_{\Omega _{kmn}^{3,bl}:\;{\rm{Interference}}\;{\rm{from}}\;{\rm{other}}\;{\rm{sensing}}\;{\rm{users}}} + u_{kmn}^{{\rm{S}},bl},
\end{align}
where ${{\rm \varphi}_{kmn}}$, $u_{kmn}^{{\rm{S}},bl}$, and $\rho _{kmn}$,   when $k\in{\mathbb K}_{\rm S}$, respectively, represent phase noise, the corresponding reception Gaussian noise with mean zero and power $\sigma_k$,  and the round-trip attenuation factor of the channel between user $k$  and its nearby target on the $(m,n)$-th RB with assuming covariance $\lambda _{kmn}={\frac{{c_o^2\delta _k^{{\rm{RCS}}}}}{{{{\left( {4\pi } \right)}^3}f_m^2r_k^4}}} $.
Here, $\delta _k^{{\rm{RCS}}}$  represents the RCS of the target around user $k$.

Then, under the assumption that ${s_{kmn}^{bl}}$ and ${s_{k'mn}^{bl}}$ for $k,k'\in{\mathbb K}_{\rm S}$ are orthogonal sequences, we know that the interference from other sensing users can be cancelled after the matched filter. Consequently, we have
\begin{align}
\bar y_{kmn}^{\rm{S}} &= \frac{1}{{BL}}\sum\nolimits_{b = 0}^{B - 1} {\sum\nolimits_{l = 0}^{L - 1} {y_{kmn}^{{\rm{S}},bl}{{\left( {\Omega_{kmn}^{1,bl}} \right)}^*}} }\nonumber\\
  &=\sqrt {{p_{kmn}}} {{\bar \rho }_{kmn}} + \bar u_{kmn}^{\rm{S}},
\end{align}
where ${\bar \rho _{kmn}} = \frac{1}{{BL}}\sum\nolimits_{b = 0}^{B - 1} {\sum\nolimits_{l = 0}^{L - 1} {\left( {{\rho _{kmn}}{e^{j{\varphi _{kmn}}}}\Omega _{kmn}^{1,bl}} \right){{\left( {\Omega _{kmn}^{1,bl}} \right)}^*}} }$ and
$\bar u_{kmn}^{\rm{S}} = \frac{1}{{BL}}\sum\nolimits_{b = 0}^{B - 1} {\sum\nolimits_{l = 0}^{L - 1} {\left( {\Omega _{kmn}^{2,bl} + u_{kmn}^{{\rm{S}},bl}} \right){{\left( {\Omega_{kmn}^{1,bl}} \right)}^*}} } $.
Moreover, upon denoting $\chi _{kk'mn}^{\rm{S}} = {\left| {{\bf{h}}_{kmn}^{\rm{H}}{\bf{w}}_{k'mn}^{{\rm{tx}}}} \right|^2}$ for $k\in{\mathbb K}_{\rm S}$ and $k'\in{\mathbb K}_{\rm C}\cup{\mathbb K}_{\rm P}$, we have
${\rm{E}}\left( {{{\left| {\bar u_{kmn}^{\rm{S}}} \right|}^2}} \right) = \frac{1}{BL}\left(I_{kmn}^{\rm{S}}+\sigma_k\right)$ where $I_{kmn}^{\rm{S}} \buildrel \Delta \over =  \sum\nolimits_{k' \in {{\mathbb K}_{\rm{C}}} \cup {{\mathbb K}_{\rm{P}}}} {{a_{k'mn}}{p_{k'mn}}} \chi _{kk'mn}^{\rm{S}}  $.

Next,  we  approximate ${{\bar \rho }_{kmn}}$ and $\bar u_{kmn}^{\rm{S}}$ as independent complex Gaussian distributions,  given by
${{\bar \rho }_{kmn}} \sim {\cal CN}\left( 0,\lambda  _{kmn}\right)$  and $\bar u_{kmn}^{\rm{S}} \sim {\cal CN}\left( 0,\frac{1}{BL}\left(I_{kmn}^{\rm{S}}+\sigma_k\right)\right)$, respectively.
Then, the hypothesis testing problem can be formulated as
\begin{align}
\bar y_{kmn}^{\rm{S}} = \left\{ {\begin{array}{*{20}{l}}
{{\cal H}_k^0:\bar u_{kmn}^{\rm{S}}},\\
{{\cal H}_k^1:\sqrt {{p_{kmn}}} {{\bar \rho }_{kmn}}  + \bar u_{kmn}^{\rm{S}}}.
\end{array}} \right.
\end{align}

Consequently, the corresponding detector is expressed as
\begin{align}
E_{kmn}^{\rm{S}} = {\left| {\bar y_{kmn}^{\rm{S}}} \right|^2}\underset{{\mathcal H}_k^1}{\overset{{\mathcal H}_k^0}{\lessgtr}} {\bar E }_k^{\rm S}  ,
 \end{align}
where $ {\bar E }_k^{\rm S}$ is the constant threshold that can control the probability of false alarm (FA).
Since both the real and imaginary parts of ${\bar y_{kmn}^{\rm{S}}}$ are independent and normally distributed with mean zero and the same variance, $E_{kmn}^{\rm{S}} $ is distributed as:
\begin{align}
E_{kmn}^{\rm{S}} \sim \left\{ {\begin{array}{*{20}{l}}
{{\cal H}_k^0:\frac{{I_{kmn}^{\rm{S}} + {\sigma _k}}}{{2BL}}{\varpi _k}},\\
{{\cal H}_k^1:\frac{1}{2}\left( {{p_{kmn}}{\lambda _{kmn}} + \frac{{I_{kmn}^{\rm{S}} + {\sigma _k}}}{{BL}}} \right){\varpi _k}},
\end{array}} \right.
 \end{align}
where ${\varpi _k}$ is a chi-square random variable with 2 degrees of freedom.

From \cite{fishler2006spatial},  the FA probability  can be given by
\begin{align}
{\cal P}_{kmn}^{{\rm{FA}}}& = \Pr \left( {E_{kmn}^{\rm{S}} > \bar E_k^{\rm{S}}\left| {{\cal H}_k^0} \right.} \right) \nonumber\\
 &= \Pr \left( {{\varpi _k} > \frac{{2BL}}{{I_{kmn}^{\rm{S}} + {\sigma _k}}}\bar E_k^{\rm{S}}} \right).
 \end{align}
Then, by fixing the  FA probability  as a required constant ${\cal P}_{k}^{{\rm{FA}}} $,  we obtain
\begin{align}\bar E_k^{\rm{S}} = \frac{{I_{kmn}^{\rm{S}} + {\sigma _k}}}{{2BL}}F_{{\varpi _k}}^{ - 1}\left(1- {{\cal P}_{k}^{{\rm{FA}}}} \right),
 \end{align}
where $F_{{\varpi _k}}$ and $F_{{\varpi _k}}^{ - 1}$ are the cumulative distribution function (CDF) of the chi-square random variable  ${\varpi _k}$ and its
inverse function, respectively. Since ${\varpi _k}$  is with 2 degrees of freedom, we have ${F_{{\varpi _k}}}\left( x \right) = \int\nolimits_0^x {\frac{1}{2}{e^{ - \frac{t}{2}}}} {d_t}=1-e^{-\frac{x}{2}}$.

Consequently, the achieved first KPI of \mbox{Type-$\rm S$} service, i.e., the detection probability, is expressed by \cite{fishler2006spatial}
\begin{align}
Q_{kmn}^{{\rm S},1} &\buildrel \Delta \over = \Pr \left( {E_{kmn}^{\rm{S}} > \bar E_k^{\rm{S}}\left| {{\cal H}_k^1} \right.} \right)\nonumber\\
   &  = 1 - {F_{{\varpi _k}}}\left( {\frac{{2\bar E_k^{\rm{S}}}}{{{p_{kmn}}{\lambda_{kmn}} + \frac{{I_{kmn}^{\rm{S}} + {\sigma _k}}}{{BL}}}}} \right)\nonumber\\
& = 1 - {F_{{\varpi _k}}}\left( {\frac{1}{{z _{kmn}^{\rm{S}} + 1}}}{{F_{{\varpi _k}}^{ - 1}\left( 1-{{\cal P}_{k}^{{\rm{FA}}}} \right)}} \right), \label{eq21}
\end{align}
where   $z _{kmn}^{\rm{S}} = \frac{{BLp_{{kmn}}{\lambda _{kmn}}}}{{I_{kmn}^{\rm{S}} + {\sigma _k}}} $ for $k\in {\mathbb K}_{\rm S} $.

Finally, we denote the latency of \mbox{Type-$\rm S$} service for user $k$ within the $(m,n)$-th RB  by $Q_{kmn}^{{\rm S},2}  \buildrel \Delta \over =  nLT$.
Then, from \eqref{eqVos},  the VoS of user $k \in {\mathbb K}_{\rm S}$ is expressed as
\begin{equation}
V_k^{\rm{S}} = \sum\nolimits_{m = 1}^M {\sum\nolimits_{n = 1}^N {{a_{kmn}}\prod\nolimits_{i = 1}^2 {{{\left( {V_{kmn}^{{\rm{S}},i}} \right)}^{w_{k}^{{\rm{S}},i}}}} } } ,\label{eqVKS}
\end{equation}
where $V_{kmn}^{{\rm{S}},i} = {\cal V}\left( {Q_{kmn}^{{\rm{S}},i},\tilde Q_{k}^{{\rm{S}},i},\alpha _k^{{\rm{S}},i},\beta _k^{{\rm{P}},i}} \right)$ is calculated by
\eqref{eqQos2}-\eqref{equationQos2b} and ${w_{k}^{{\rm{S}},i}}$ is defined after \eqref{eqVos}.

\section{Problem Formulation and Transformation}\label{Sec4}
This section mathematically formulates the  VoS maximization problem and transforms it into an equivalent but tractable formulation to facilitate algorithm development.

\subsection{Problem Formulation}

Given that sensing users typically have limited maximum transmission power, we assume that each sensing user operates at its maximum available power during the sensing process.
Consequently, we attempt to jointly optimize power allocation at the BS  and the MDMA assignment for all users on all RBs, thus maximizing the proportional VoS of all concurrent heterogeneous services. Mathematically, upon denoting the MDMA assignment by  ${\bf{a}} = {[{\bf{a}}_1^{\rm T},\cdots,{\bf{a}}_N^{\rm T}]^{\rm T}}\in {{\mathbb C}^{KMN\times1}}$ with ${{\bf{a}}_n} = {[{a_{11n}},\cdots,{a_{kmn}},\cdots, {a_{KMn}}]^{\rm T}}\in {{\mathbb C}^{KM\times1}}$ and
 the power allocation by  ${\bf{p}} = {[{\bf{p}}_1^{\rm T},\cdots,{\bf{p}}_N^{\rm T}]^{\rm T}}\in {{\mathbb C}^{({\left| {{{\mathbb K}_{\rm C}}} \right| + \left| {{{\mathbb K}_{\rm P}}} \right|})MN\times1}}$ with ${{\bf{p}}_n} = {[{p_{11n}},,\cdots,{p_{kmn}},,\cdots, {p_{({\left| {{{\mathbb K}_{\rm C}}} \right| + \left| {{{\mathbb K}_{\rm P}}} \right|})Mn}}]^{\rm T}}\in {{\mathbb C}^{({\left| {{{\mathbb K}_{\rm C}}} \right| + \left| {{{\mathbb K}_{\rm P}}} \right|})M\times1}}$, the optimization problem, considering proportional fairness, is formulated as follows:
\begin{subequations}
\begin{align}
\label{PF}
\mathop {\max }\limits_{{\bf{a}},{\bf{p}}} & \prod\limits_{{\rm{X}} \in \left\{ {{\rm{C,P,S}}} \right\}} {\prod\limits_{k \in {{\mathbb K}_{\rm{X}}}} {V_k^{\rm X}} }  \tag{\bf P1}\\
{\rm{s}}{\rm{.t}}{\rm{.}}\;&\sum\nolimits_{k \in {{\mathbb K}_{\rm{C}}} \cup {{\mathbb K}_{\rm{P}}}} {\sum\nolimits_{m = 1}^M {{p_{kmn}}}  \le P_{\rm max}},\forall n,\label{PFCA} \\
&\sum\nolimits_{k \in {{\mathbb K}}} {{a_{kmn}}}   \le A_{\rm max}, \forall m, \forall n,\label{PFCB}\\
&\sum\nolimits_{m = 1}^M {\sum\nolimits_{n = 1}^N {{a_{kmn}}} }  = 1, \forall k, \label{PFCC}\\
&{p_{qmn}}\chi _{kqmn}^{\rm{C}} \ge {p_{jmn}}\chi _{kjmn}^{\rm{C}}, \forall m,\forall n, 1\le k\le \left| {{{\mathbb K}_{\rm{C}}}} \right|, \nonumber\\
&\;\;\;{\rm{if}} \; a_{k}^{mn}\!\!=a_{j}^{mn}\!=\!a_{q}^{mn}=1\&1\le j<q\le\left| {{{\mathbb K}_{\rm{C}}}} \right|,\label{PFCF}\\
&0\le{p_{kmn}} \le {a_{kmn}}P_{\rm max},  \forall m,\forall n, {\forall k \in {{\mathbb K}_{\rm{C}}} \cup {{\mathbb K}_{\rm{P}}}},\label{PFCE}\\
&{a_{kmn}} \in \left\{ {0,1} \right\},\forall k,\forall m,\forall n,\label{PFCD}
  \end{align}
\end{subequations}
where constraint \eqref{PFCA} implies that the maximum transmission power should be smaller than $P_{\rm max}$ at time sub-frame $n$;
constraints \eqref{PFCB} and \eqref{PFCC} ensure that one RB can accommodate a maximum of $A_{\rm max} $ services, and each service is accessed only once across all RBs; constraint \eqref{PFCF} is for power allocation fairness in NOMA;
constraints \eqref{PFCE} and \eqref{PFCD}  represent practical power allocation and integer MDMA assignment conditions, respectively, which ensure the feasibility and relevance of the problem formulation.

\subsection{Problem Transformation}

Due to the integer constraint \eqref{PFCD},  \ref{PF} belongs to the MINLP family. Thus, it is challenging to find the optimal solution. Hence, we transform  \ref{PF} into an equivalent but tractable formulation to facilitate algorithm development.

To begin with, we examine the monotonicity and curvature properties of the functions associated with problem~\ref{PF} in the following lemmas.
\begin{lemma}\label{lemmaA}
For $z_{kmn}^{\rm  X}>0 $, we have the following  results:
\begin{itemize}
\item  For \mbox{Type-$\rm C$} services, the first HKPI $Q_{kmn}^{{\rm{C}},1}$ is concave and nondecreasing in $z_{kmn}^{\rm  C}>0 $.
\item For Type-$\rm P$ services, the first three LKPIs $Q_{kmn}^{{\rm{P}},i}$ with $1\le i\le 3$ are convex and nonincreasing in $z_{kmn}^{\rm  P}>0 $.
\item For \mbox{Type-$\rm S$} services, the first HKPI $Q_{kmn}^{{\rm{S}},1}$ is  nondecreasing $z_{kmn}^{\rm  P}>0 $. Additionally, it is concave in  $z_{kmn}^{\rm  P}>0 $ with  the further assumption that
     $\frac{1}{e^2}\le {\cal P}_{k}^{{\rm{FA}}}$.
\end{itemize}
\end{lemma}
\begin{IEEEproof} Please refer to Appendix~\ref{appendix2}.
\end{IEEEproof}
\begin{lemma}\label{lemmaB} For elastic parameters $\alpha>0$ and $0<\beta< 1$,  ${\log {\cal V}\left( {Q,\tilde Q,\alpha ,\beta } \right)}$ exhibits the following properties:
\begin{itemize}
\item Under HKPI, it is nondecreasing over $Q \in (0, +\infty)$ and concave over $Q \in (\beta \tilde{Q},+\infty)$.
\item Under LKPI, it is nonincreasing over $Q \in (0, +\infty)$ and concave over $Q \in (0, \tilde{Q}/\beta)$.
\end{itemize}
\end{lemma}
\begin{IEEEproof}
Please refer to Appendix~\ref{appendix3}.
\end{IEEEproof}

Besides, based on the definitions of $V_k^{\rm{X}}$ in \eqref{eqVKC}, \eqref{eqVKP}, and \eqref{eqVKS}, maximizing the objective of \ref{PF} is equivalent to maximizing the following expression:
\begin{align}
&\log \left( {\prod\limits_{{\rm{X}} \in \left\{ {{\rm{C}},{\rm{P}},{\rm{S}}} \right\}} {\prod\limits_{k \in {{\mathbb K}_{\rm{X}}}} {V_k^{\rm{X}}} } } \right) \nonumber\\
 =& \sum\limits_{n = 1}^N {\underbrace {\sum\limits_{m = 1}^M {\sum\limits_{{\rm{X}} \in \left\{ {{\rm{C}},{\rm{P}},{\rm{S}}} \right\}} {\sum\limits_{k \in {{\mathbb K}_{\rm{X}}}} {\left[ {{a_{kmn}}\sum\limits_{i = 1}^{\left| {{{\mathbb Q}_{\rm{X}}}} \right|} {w_k^{{\rm{X}},i}} \log V_{kmn}^{{\rm{X}},i}} \right]} } } }_{{{\cal L}_n}\left( {{{\bf{a}}_n},{{\bf{z}}_n}} \right)}}.
\end{align}
where $\mathcal{L}_n \left( \mathbf{a}_n, \mathbf{z}_n \right)$ denotes the logarithmic form of the overall proportional VoS $V_k^{\rm{X}}$ of all services in the $n$-th time subframe. Here,
 ${\bf{z}}_n^{\rm T}\in {{\mathbb C}^{KM\times1}}$ represents the auxiliary  variable, which consists of elements from $\left\{z_{kmn}^{\rm X}, \; \forall m, k \in {{\mathbb K}_{\rm{X}}},{\rm{X}} \in \left\{ {{\rm{C}},{\rm{P}},{\rm{S}}} \right\} \right\}$.

Next, upon denoting ${\bf{z}}={\left[ {{\bf{z}}_1^{\rm T},\ldots,{\bf{z}}_N^{\rm T}} \right]^{\rm T}} \in {{\mathbb C}^{KMN\times1}} $ and based on the monotonicity properties analyzed in Lemmas \ref{lemmaA} and \ref{lemmaB},  problem \ref{PF} can be equivalently re-expressed as:
\begin{subequations}
\begin{align}
\mathop {\max }\limits_{{\bf{a}},{\bf{p}},{\bf{z}}} \;& {\cal L}\left({\bf a}, {\bf z}\right)=\sum\nolimits_{n = 1}^N {{\cal L}_n }\left({\bf a}_n,{\bf z}_n\right) ,  \tag{\bf P2} \label{P2}\\
{\rm{s}}{\rm{.t}}{\rm{.}}\;& z_{kmn}^{
\rm C} \le \frac{{{p_{kmn}}\chi _{qkmn}^{\rm{C}}}}{{\sum\nolimits_{j = 1}^{k - 1} {{a_{jmn}}{p_{jmn}}\chi _{qjmn}^{\rm{C}}}  + {\sigma _q}}},\nonumber\\
&\qquad\quad  {\rm if}\;{a_{qmn}} =1, 1 \le q \le k \le \left| {{{\mathbb K}_{\rm{C}}}} \right|, \forall m,\forall n,\label{eqp2c1}\\
&z_{kmn}^{\rm{P}} \le \frac{ \!\sum\nolimits_{k' \in {{\mathbb K}_{\rm{C}}} \cup {{\mathbb K}_{\rm{P}}}}\!\!  {a_{k'mn}}{p_{k'mn}}\chi _{kk'mn}^{\rm{P}}}{{2{\sigma _0}}},\nonumber\\
&\qquad\qquad\qquad\qquad\qquad\quad\quad k \in {{\mathbb K}_{\rm{P}}}, \forall m,\forall n,\label{eqp2c2}\\
&z_{kmn}^{\rm{S}} \le \frac{{BL{p_{kmn}}{\lambda _{kmn}}}}{\!{\sum\nolimits_{k' \in {{\mathbb K}_{\rm{C}}} \cup {{\mathbb K}_{\rm{P}}}} \!\!{{a_{k'mn}}{p_{k'mn}}} \chi _{kk'mn}^{\rm{S}} + {\sigma _k}}},\nonumber\\
&\qquad\qquad\qquad\qquad\qquad\qquad\quad k \in {{\mathbb K}_{\rm{S}}}, \forall m,\forall n, \label{eqp2c3}\\
&z_{kmn}^{\rm  X}\ge0, k \in {{\mathbb K}_{\rm{X}}},\;  {\rm{X}} \in \left\{ {{\rm{C}},{\rm{P}},{\rm{S}}} \right\},  \forall m,n, \label{eqp2c4} \\
&\eqref{PFCA}-\eqref{PFCD}.
\end{align}
\end{subequations}

Then, we examine the monotonicity and curvature properties of the objective function of problem \ref{P2} in the following theorem to facilitate the algorithm development.
\begin{theorem}\label{lemma4}
 Given ${\bf a}_n$, function ${\cal L}_n\left({\bf a}_n,{\bf z}_n\right)$ is  non-decreasing for ${\bf z}_n \ge { \bf 0}$. Furthermore, it is concave for ${\bf z}_n \ge  {{\bf{z}}_n^{{\rm{min}}}}$ under the assumption that $\frac{1}{e^2} \leq {\cal P}_k^{\rm{FA}}$ for  $k\in{\mathbb K}_{\rm S}$, where the $km$-th element of the vector ${{\bf{z}}_n^{{\rm{min}}}}$ is given by
 \begin{subequations}
 \begin{align}
{\left[ {{\bf{z}}_n^{{\rm{min}}}} \right]_{km}} &= {2^{\beta _k^{{\rm C},1}\tilde Q_k^{{\rm C},1}}} - 1,  {\rm{for}}\;1\le k \le \left| {{{\mathbb K}_{\rm{C}}}} \right|,\\
{\left[ {{\bf{z}}_n^{{\rm{min}}}} \right]_{km}} &= \max \left\{ {\frac{{\beta _k^{{\rm{P}},1}I_{kmn}^{{\rm{P,}}\theta }}}{{\tilde Q_k^{{\rm{P}},1}}},\frac{{\beta _k^{{\rm{P}},2}I_{kmn}^{{\rm{P,}}d}}}{{\tilde Q_k^{{\rm{P}},2}}},\frac{{\beta _k^{{\rm{P}},3}I_{kmn}^{{\rm{P,}}v}}}{{\tilde Q_k^{{\rm{P}},3}}}} \right\},\nonumber\\
&\qquad\qquad\quad{\rm{for}}\;\left| {{{\mathbb K}_{\rm{C}}}} \right| < k \le \left| {{{\mathbb K}_{\rm{C}}}} \right| + \left| {{{\mathbb K}_{\rm{P}}}} \right|,\\
{\left[ {{\bf{z}}_n^{{\rm{min}}}} \right]_{km}}& = \frac{{F_{{\varpi _k}}^{ - 1}\left( 1-{{\cal P}_k^{{\rm{FA}}}} \right)}}{{F_{{\varpi _k}}^{ - 1}\left( {1 - \beta _k^{{\rm{S}},1}\tilde Q_k^{{\rm{S}},1}} \right)}} - 1,\nonumber\\
&\qquad\qquad\qquad\quad{\rm{for}}\;\left| {{{\mathbb K}_{\rm{C}}}} \right| + \left| {{{\mathbb K}_{\rm{P}}}} \right| < k \le K,
\end{align}
\end{subequations}
\end{theorem}
if   $a_{kmn}=1$; otherwise  ${\left[ {{\bf{z}}_n^{{\rm{min}}}} \right]_{km}}=0$.
\begin{IEEEproof}
Please refer to Appendix~\ref{appendix4}
\end{IEEEproof}

However,  \ref{P2} remains non-convex and belongs to MINLP \cite{chen2024semantic,shan2024resource} due to the integer constraint \eqref{PFCD} and the non-convex constraints \eqref{eqp2c1} and \eqref{eqp2c3}, thus requiring an exhaustive search for optimal solutions, which is computationally impractical.
Thanks to Theorem \ref{lemma4}, when $\bf a$ is fixed,  \ref{P2} reduces to a sub-problem with a concave and non-decreasing objective function of $\bf z$.
This serves as the foundation for developing efficient solutions in the following sections.

\section{Optimal Solution: Monotonic Optimization-Aided Dynamic Programming (MODP)} \label{Sec5}
In this section, we propose the MODP algorithm to find the optimal solution for Problem \ref{P2}. This algorithm employs a DP framework to recursively determine the optimal MDMA assignment. Within each recursive step, an MO algorithm is applied to solve the corresponding power allocation sub-problem.

\subsection{DP Recursion Framework}
To transform \ref{P2} into a DP recursion framework, we define the service assignment state at the first $n$ time sub-frames in the DP recursion framework as
\begin{align}
{{\mathbb S}_n} = \bigcup\nolimits_{i = 1}^n  {{{\mathbb A}_i}}, \label{eqstate}
\end{align}
where ${{\mathbb S}_n}$ is  the index set of services assigned on the first $n$ time sub-frames across all sub-bands and
${\mathbb A}{_i} = \left\{ {k\left| {{a_{kmi}} = 1,\forall k \in \bar {\mathbb S}_{{i - 1}},\forall m} \right.} \right\} \subseteq {\mathbb K}$ is the index set of services assigned on the $i$-th time sub-frame across all sub-bands.
Here, $\bar {\mathbb S}_{{i }}$ is the complement set of $ {\mathbb S}_{{i }}$, which satisfies
${\mathbb S}{_i} \cap \bar {\mathbb S}{_i} = \emptyset$ and ${\mathbb S}{_i} \cup \bar {\mathbb S}{_i} = {\mathbb K}$.

Then, we can observe that the initial service assignment and the final state must be given by ${\mathbb S}_0=\emptyset$ and ${\mathbb S}_N={\mathbb K}$.
Besides, from \eqref{eqstate}, we have the following state transition formula:
\begin{align}
{{\mathbb S}_n} = {{\mathbb S}_{n-1}}\bigcup {{\mathbb A}_n}. \label{eqstatemarkov}
\end{align}

\begin{algorithm}[t]
\caption{Optimal MODP Algorithm for \ref{P2}} \label{DPMPAlgorithm}
 {
Input ${\mathbb S}_0= \emptyset$, ${\mathbb S}_N= {\mathbb K}$, and ${{\cal U}^\star}({{\mathbb S}_{0}})=0$\;
    \For{time sub-frame $n=1, \dots, N$}{
        \For{ each state ${\mathbb S}_n$ on  time sub-frame $n$}{
            \For{all possible  ${\mathbb S}_{n-1}$ on subcarrier $n-1$}{
                \For {all possible ${{\bf{a}}_n}$ given  ${\mathbb S}_n$ and  ${\mathbb S}_{n-1}$}{
                Calculate $
{{{\cal U}^\star}\left( {{{\bf{a}}_n}\left| {{{\mathbb S}_{n - 1}},{{\mathbb S}_n}} \right.} \right)}$ by solving  \ref{P3a} using MO in Algorithm 2
            }
        Calculate $\Delta {^\star}{\cal U}\left( {{{\mathbb S}_{n - 1}},{{\mathbb S}_n}} \right)$ by solving  \ref{P3}\;
        }

        Calculate ${{\cal U}^\star}({{\mathbb S}_n}) = \mathop {\max }\limits_{\forall {{\mathbb S}_{n - 1}}} \left[ {\Delta ^{\star}{\cal U}\left( {{{\mathbb S}_{n - 1}},{{\mathbb S}_n}} \right) + {{\cal U}^\star}({{\mathbb S}_{n - 1}})} \right]$

    }
}

 Recover the optimal  $\bf a$, $\bf p$, and $\bf z$ by backtracking on the state transition path, one by one, from ${\mathbb S}_N$ to ${\mathbb S}_0$ that can achieve the maximum ${{\cal U}^\star}({{\mathbb S}_N})$\;
  Output the optimal  $\bf a$, $\bf p$, $\bf z$ and the utility ${{\cal U}^\star}({{\mathbb S}_N})$.}
\end{algorithm}

This implies that once ${{\mathbb S}_{n-1}}$  is given, the state transition from ${{\mathbb S}_{n-1}}$ to ${{\mathbb S}_{n}}$ depends solely on the previous state ${{\mathbb S}_{n-1}}$ and is independent of ${{\mathbb S}_{n-2}}$. Thus, the total proportional VoS of the first $n$ time sub-frames can be rewritten as
\begin{align}
 \sum\limits_{i = 1}^n {\cal L}_i\left({\bf a}_i,{\bf z}_i\right) = \sum\limits_{i = 1}^{n-1}{\cal L}_i\left({\bf a}_i,{\bf z}_i\right)  +\underbrace {{\cal L}_n\left({\bf a}_n,{\bf z}_n\right)}_{\Delta {\cal U}\left( {{{\mathbb S}_{n - 1}},{{\mathbb S}_n}} \right)}.
\end{align}
Here,  ${\cal L}_n\left({\bf a}_n,{\bf z}_n\right)$ is represented by $\Delta {\cal U}\left( {{{\mathbb S}_{n - 1}},{{\mathbb S}_n}} \right) $, which denotes the proportional VoS achieved on the $n$-th time sub-frame across all sub-bands when the state transitions from ${{\mathbb S}_{n - 1}}$ to ${{\mathbb S}_{n }}$.

Next, let ${\cal U}^{\star}\left( {{{\mathbb S}_n}} \right)$ be the optimal proportional VoS achieved  over the first $n$ time sub-frames across all sub-bands when the state transitions to ${{\mathbb S}_{n }}$. Then, problem \ref{P2} can be transformed into the following DP recursion framework, i.e.,
\begin{align}
{{\cal U}^\star}({{\mathbb S}_n}) = &\mathop {\max }\limits_{\forall {{\mathbb S}_{n - 1}}} \left[ {\Delta ^{\star}{\cal U}\left( {{{\mathbb S}_{n - 1}},{{\mathbb S}_n}} \right) + {{\cal U}^\star}({{\mathbb S}_{n - 1}})} \right],\nonumber\\
&\qquad\qquad\qquad\qquad\qquad\qquad1 \le n \le N, \label{eqDP1}
\end{align}
where ${{\cal U}^\star}({{\mathbb S}_0})=0$, and $\Delta ^{\star}{\cal U}\left( {{{\mathbb S}_{n - 1}},{{\mathbb S}_n}} \right)$ is the optimal proportional VoS achieved on the $n$-th time sub-frame given states ${{\mathbb S}_{n-1}}$  and ${{\mathbb S}_n}$.
Besides, if ${{\mathbb S}_{n-1}}$  and ${{\mathbb S}_n}$ are  given, we know the services assigned in the $n$-th time sub-frame, i.e., ${\mathbb A}{_n}={{\mathbb S}_{n}}\backslash {{\mathbb S}_{n - 1}}$. In other words, in the $n$-th time sub-frame, given ${\mathbb A}{_n}$,  ${\bf  a}_n$ must satisfy the following condition:
\begin{align}
\sum\nolimits_{m = 1}^M {{a_{kmn}}}  = \left\{ {\begin{array}{*{20}{l}}
{1,\;\;{\rm{if}}\;k \in {{\mathbb A}_n},}\\
{0,\;\;{\rm{otherwise.}}}
\end{array}} \right. \label{eqakmn31}
\end{align}
Moreover,  $ \Delta ^{\star}{\cal U}\left( {{{\mathbb S}_{n - 1}},{{\mathbb S}_n}} \right) $ can be obtained by exhaustively searching for the maximum achievable value across all possible allocations of the services in ${\mathbb A}{_n}$ across $M$ sub-bands.
 Mathematically, we need to solve the following sub-problem
\begin{align}
\Delta {^\star}{\cal U}\left( {{{\mathbb S}_{n - 1}},{{\mathbb S}_n}} \right) = \mathop {\max }\limits_{{{\bf{a}}_n}} &\;\left\{ {{{\cal U}^\star}\left( {{{\bf{a}}_n}\left| {{{\mathbb S}_{n - 1}},{{\mathbb S}_n}} \right.} \right)} \right\}, \tag{\bf P3} \label{P3}\\
{\rm s.t.}\; &\; \eqref{eqakmn31}, \nonumber
\end{align}
where  ${{{\cal U}^\star}\left( {{{\bf{a}}_n}\left| {{{\mathbb S}_{n - 1}},{{\mathbb S}_n}} \right.} \right)}$ represents the optimal value obtained by assigning services in ${\mathbb A}{_n}$ across $M$ sub-bands for a given allocation  ${\bf a}_n$ satisfying  \eqref{eqakmn31}, i.e.,
\begin{align}
{{{\cal U}^\star}\left( {{{\bf{a}}_n}\left| {{{\mathbb S}_{n - 1}},{{\mathbb S}_n}} \right.} \right)} = \mathop {\max }\limits_{{{\bf{p}}_n},{{\bf{z}}_n}} &\;{{\cal L}_n} \left({\bf a}_n,{\bf z}_n\right) \tag{\bf P3a} \label{P3a}\\
{\rm{s}}{\rm{.t}}{\rm{.}}\;&\; \eqref{PFCA}, \eqref{PFCF},\eqref{PFCE},  \eqref{eqp2c1}-\eqref{eqp2c4}.\nonumber
\end{align}

If the optimal solution to  \ref{P3a} is obtained, the optimal solution to  \ref{P3} can be determined by comparing all potential MDMA assignments in the $n$-th time frame, i.e., ${\bf a}_n$, under constraint \eqref{eqakmn31}. Consequently, the optimal solution to problem \ref{P2} can be derived by recursively computing \eqref{eqDP1}.
Finally, after obtaining ${{\cal U}^\star}({{\mathbb S}_N})$ defined in \eqref{eqDP1}, we can recover the optimal $\bf a$, $\bf p$, and $\bf z$, by backtracking through the state transition path. These detailed procedures are summarized in Algorithm~\ref{DPMPAlgorithm}.
Therefore, the remaining challenge is to obtain the optimal solution to the non-convex problem \ref{P3a}.

\subsection{Optimal Solution to Problem \ref{P3a}  }
In this part, we utilize the MO technique to find the optimal solution to problem \ref{P3a}.
Specifically, we first introduce the preliminaries of the MO technique \cite{chen2025knowledge,chen2024otfs,chen2025radiation}.
\begin{itemize}
\item {\bf Box}: For ${\bf z} \in {\mathbb{R}}_{+}^{D}$, the  $D$-dimensional box with   vertex ${\bf z}$ represents
 the hyperrectangle $\left[ {\bf 0}, {\bf z} \right] = \left\{ {\bf x} \mid {\bf 0} \leq {\bf x} \leq {\bf z} \right\}$.

\item {\bf Normal}: An infinite set ${\cal Z} \subset {\mathbb R}_{+}^{D}$ is normal if, for every ${\bf z} \in {\cal Z}$, the box $\left[ {\bf 0}, {\bf z} \right]$ is fully  contained within ${\cal Z}$.

\item {\bf Polyblock}: Let ${\mathbb V}$ be a finite set of vertices.
The corresponding polyblock ${\cal B}({\mathbb V} )$  is defined as  the union of all boxes $\left[ {\bf 0}, {\bf z} \right]$ associated with each vertex ${\bf z} \in {\mathbb V} $, i.e.,
  \begin{align}
{\cal B}({\mathbb V}) = \mathop  \cup \limits_{{\bf{z}} \in {\mathbb V}} \left[ {{\bf 0},{\bf{z}}} \right],
\end{align}
where  ${\mathbb V}$ serves as the vertex set of this polyblock.
\item {\bf Projection}: Given a non-empty normal set ${\cal Z} \subset {\mathbb R}_{+}^{D}$ and a vertex ${\bf z}$, the projection of ${\bf z}$ onto the boundary of ${\cal Z}$ is denoted as ${\rm Proj}({\bf z}) = \delta^\star{\bf z}$, where $\delta^\star$ is defined as
${\delta ^\star} = \max \left\{ \delta \left| \delta  {{\bf{z}} } \in {\cal Z},0 \le \delta  \le 1 \right. \right\}$.
\item {\bf MO problem}: A problem belongs to the MO family if it can be expressed as
   \begin{align}
\mathop {\max }\limits_{\bf{z}}\;\; {\cal W}\left( {\bf{z}} \right), \quad {
\rm s.t.}\; {\bf{z}} \in {\cal Z}, \label{MO}
\end{align}
where $ {\cal W}\left( {\bf{z}} \right)$  is an increasing function of $\bf z$  and $ {\cal Z}$ is a non-empty normal closed set.
\end{itemize}

Based on the above preliminaries of MO and Theorem \ref{lemma4},  we know that \ref{P3a} is an MO problem. Consequently, its optimal solution can be obtained using the polyblock algorithm, which is an efficient method for solving MO problems.

Before introducing the algorithm details, we first outline the principle of the polyblock algorithm. Firstly, since the objective function increases monotonically with $\bf{z}$, the optimal $\bf{z}$ must reside on the boundary of the feasible region ${\cal Z}$. Then, as shown in Fig. \ref{fig2},  we iteratively apply polyblocks to approximate this boundary that includes the optimal solution.
In each iteration, the polyblock space is refined by subdividing it and eliminating regions that either lie outside the feasible set or cannot possibly contain the optimal solution. This process progressively narrows the gap between the current boundary and the outer limit of the polyblock space, thereby reducing the region where the optimal solution may reside.
Finally, the polyblock can approach the boundary infinitely and find the optimal point.
The key steps of the algorithm are as follows:

\begin{figure*}[ht]
\centering
\includegraphics[width = 0.98\textwidth]{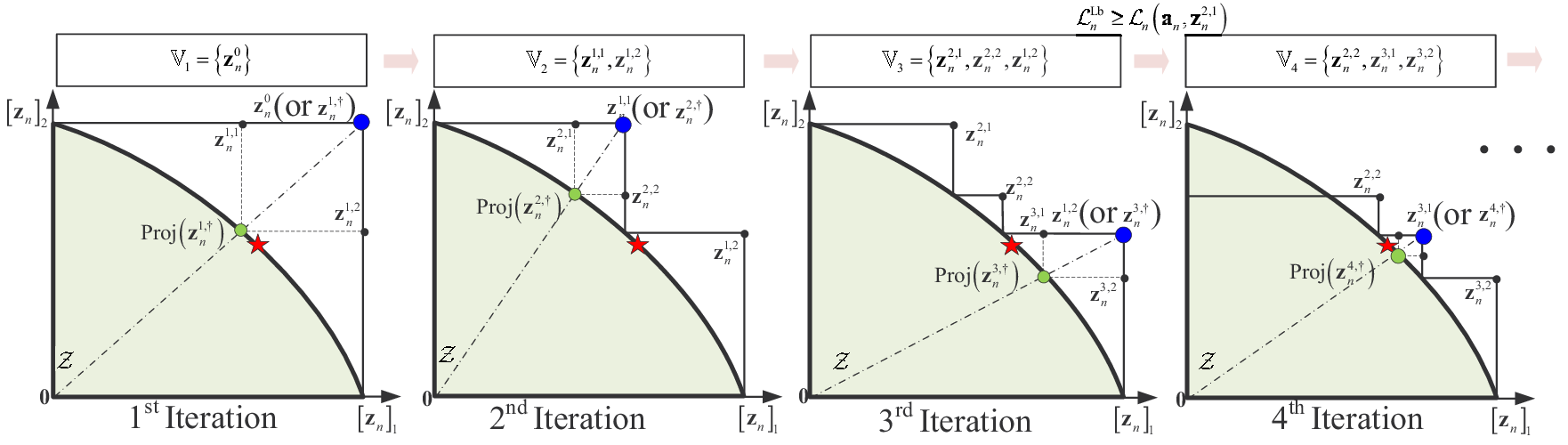}
\caption{An illustration of the polyblock algorithm when $\mathbf{z}_n\in{\mathbb R}_+^2$: the red star denotes the global optimal solution ${\bf z}_n^\star$, the blue circle represents the optimal vertex in each iteration, and the green circle indicates the corresponding projected point on the boundary.} \label{fig2}
\end{figure*}

\subsubsection{\bf Initialization} In problem \ref{P3a}, we know ${\bf z}_n\in{\mathbb R}_+^D$ where $D=KM$. Then, we can define an initial polyblock ${\cal B}\left( {{\mathbb V}_1} \right)$  enclosing the entire feasible region $\cal Z$. Specifically,
 Given \eqref{PFCA}, \eqref{eqp2c1}-\eqref{eqp2c3} with a specified ${\bf a}_n$,  the initial vertex set can be defined by ${{\mathbb V}_1} = \left\{ {{\bf{z}}_n^0} \right\}$, where the $km$-th element of the vector ${{\bf{z}}_n^0}$ is given by
 \begin{subequations}\begin{align}
\!\! {\left[ {{\bf{z}}_n^0} \right]_{km}} &=\! {P_{\max }}\mathop {\max }\limits_{\scriptstyle1 \le q \le k\hfill\atop
\scriptstyle{a_{qmn}} = 1\hfill} \frac{{\chi _{qkmn}^{\rm{C}}}}{{{\sigma _k}}},{\rm{for}}\; 1\le k \le \left| {{{\mathbb K}_{\rm{C}}}} \right|,\\
\!\! {\left[ {{\bf{z}}_n^0} \right]_{km}} &=\! \frac{{\sum\nolimits_{k' \in {{\mathbb K}_{\rm{C}}} \cup {{\mathbb K}_{\rm{P}}}}  {a_{k'mn}}{P_{\max }}\chi _{kk'mn}^{\rm{P}}}}{{2{\sigma _0}}},\nonumber\\
&\qquad\qquad\quad\quad{\rm{for}}\;\left| {{{\mathbb K}_{\rm{C}}}} \right| < k \le \left| {{{\mathbb K}_{\rm{C}}}} \right| + \left| {{{\mathbb K}_{\rm{P}}}} \right|,\\
\!\! {\left[ {{\bf{z}}_n^0} \right]_{km}}& =\! \frac{{ BL{p_{kmn}}{\lambda _{kmn}}}}{{{\sigma _k}}},\;{\rm{for}}\;\left| {{{\mathbb K}_{\rm{C}}}} \right| + \left| {{{\mathbb K}_{\rm{P}}}} \right| < k \le K,
\end{align}\end{subequations}
if   $a_{kmn}=1$; otherwise  ${\left[ {{\bf{z}}_n^0} \right]_{km}} =0$.

\subsubsection{\bf  Maximum Vertex Determination}
Upon denoting the  vertex set in the $j$-th iteration of polyblock algorithm by ${\mathbb V}_j$, we find  the vertex  belonging to ${\mathbb V}_j$
 that achieves the maximum objective function value, i.e.,
\begin{align}
{\bf z}_{n}^{j,\dag}=\arg \mathop {\max }\limits_{{\bf{z}}_n \in {\mathbb V}_{j}} {\cal L}_n\left({\bf a}_n, {\bf z}_n\right).
 \end{align}

\subsubsection{\bf  Polyblock Reduction by Projection and Bounding}
The infeasible vertex ${\bf z}_{n}^{j,\dag}$ can be projected onto the feasible boundary $\cal Z$, with the corresponding projected vertex denoted by ${\rm Proj}\left({\bf z}_{n}^\star\right)$. Then, upon utilizing this projection, a smaller polyblock can be constructed by removing ${\bf z}_{n}^{j,\dag}$ from ${\mathbb V}_{j}$ and adding the following maximum $D$ new smaller vertices to ${\mathbb V}_{j}$:
\begin{align}
 {{\mathbb V}_{j + 1}} = \left( {{{\mathbb V}_j}\backslash \left\{ {{\bf{z}}_n^{j,\dag} } \right\}} \right) \cup \left\{ {{\bf{z}}_n^{j,1},{\bf{z}}_n^{j,2}, \cdots ,{\bf{z}}_n^{j,D}} \right\},\label{eqVall}
\end{align}
where ${\bf{z}}_n^{j,d}$ is the $d$-th new added vertex, i.e.,
 \begin{align}
 {\bf{z}}_n^{j,d} = {\bf{z}}_n^ {j,\dag} - {\left[ {{\bf{z}}_n^{j,\dag} - {\rm{Proj}}\left( {{\bf{z}}_n^{j, \dag} } \right)} \right]_d} {{\bf{e}}_d}, 1 \le d \le D
 ,\label{eqproj1}
\end{align}
where $\left[ {{\bf{z}}_n^{j,\dag} - {\rm{Proj}}\left( {{\bf{z}}_n^{j,\dag} } \right)} \right]_d$ is the $d$-th element of ${{\bf{z}}_n^ {j,\dag} - {\rm{Proj}}\left( {{\bf{z}}_n^{j,\dag} } \right)} $ and ${\bf e}_d\in{\mathbb C}^{D\times1}$  is a unit vector with the $d$-th element being one and other elements being zero. Note that if the $d$-th element $ {\bf{z}}_n^{j,\dag}$ is very small, we do not need to add
 ${\bf{z}}_n^{j,d} $ into the vertex set for a better convergence.

As the iterations progress, the  optimal solution denoted by ${\bf z}_n^\star$ stays within the updated and iteratively shrinking polyblock defined by the vertex set ${\mathbb V}_j$, i.e.,
\begin{align}
{\cal B}\left( {{{\mathbb V}_0}} \right) \supset {\cal B}\left( {{{\mathbb V}_1}} \right) \cdots\supset {\cal B}\left( {{{\mathbb V}_i}} \right)\cdots \supset  {\cal B}\left(\left\{{\bf z}_n^\star\right\}\right).
\end{align}

It can be observed that the upper bound at each iteration can be defined using the maximal infeasible vertex ${\bf{z}}_n^ {j,\dag} $, i.e., ${\cal L}_n^{{\rm{Ub}}} = {{\cal L}_n}\left( {{{\bf{a}}_n},{\bf{z}}_n^{j,\dag}} \right)$.
Then, we denote the lower bound as the maximum value achieved during all iterations of the feasible projected vertex, i.e., ${\cal L}_n^{{\rm{Lb}}} \leftarrow \max \left\{ {{{\cal L}_n}\left( {{{\bf{a}}_n},{\rm{Proj}}\left( {{\bf{z}}_n^j} \right)} \right),{\cal L}_n^{{\rm{Lb}}}} \right\}$.
It is straightforward to know that the lower bound increases and the upper bound decreases as the polyblock shrinks. Finally,  the iteration  converges when ${\cal L}_n^{{\rm{Ub}}} < \left( {1 + \epsilon } \right){\cal L}_n^{{\rm{Lb}}}$. Here, $\epsilon$ is a small positive constant, which denotes the accuracy requirement of the optimal solution.

Moreover, the remaining task is to find the projected vertex ${\rm{Proj}}\left( {\bf{z}}_n^{j,\dag} \right)$. This can be solved by employing a bisection method \cite{chen2025radiation} to iteratively search for $\delta^\star$ within the initial range of $\delta^{\rm lb} = 0$ to $\delta^{\rm ub} = 1$, checking whether a feasible ${\bf p}$ can be found to make problem \ref{P3a} feasible by setting ${\bf z}_n = \delta^\star {\bf z}_n^\dag$ with $\delta^\star = \frac{{\delta^{\rm lb} + \delta^{\rm ub}}}{2}$. If a feasible solution is found, update ${\delta^{\rm lb}} \leftarrow \delta^\star$; otherwise, update ${\delta^{\rm ub}} \leftarrow \delta^\star$. Specifically, the feasibility check is a convex problem because ${\bf z}_n$ is constant in \eqref{eqp2c1} and \eqref{eqp2c3}. Thus, it can be efficiently solved using the Matlab toolbox CVX within a polynomial
complexity.

\begin{algorithm} [t]
\caption{Polyblock algorithm for \ref{P3a}}\label{algorithmMO}
 \renewcommand{\baselinestretch}{0.95}
 {{\bf INITIALIZATION: (step 2 - step 4)} \\
Initialize iteration index $i=1$, polyblock ${\cal B}(  {\mathbb V}_1)$ with ${\mathbb V}_1=\left\{{\bf z}_n^{0}\right\}$ \;
Set the lower bound, upper bound, error tolerance by ${\cal L}_n^{\rm Lb}=-{\rm Inf}$, ${\cal L}_n^{\rm Ub}={\cal L}_n({\bf a}_n, {\bf z}_n^{0})$, $\tilde \rho_1=0.05$, respectively\;
\While{${\cal L}_n^{\rm Ub}>(1+\epsilon  ){\cal L}_n^{\rm Lb} $}{

{\bf Maximum Vertex Determination:} \\

Find  the vertex in ${\mathbb V}_{j}$  that maximizes the objective function:
\begin{align}
{\bf z}_{n}^{j,\dag}=\arg \mathop {\max }\nolimits_{{\bf{z}}_n \in {\mathbb V}_{j}} {\cal L}_n\left({\bf a}_n, {\bf z}_n\right) \nonumber
 \end{align}

{\bf   Projection:} \\

Given ${\bf z} = {\bf z}_{n}^{j,\dag}$, solve projection problem  and obtain ${\rm Proj}( {\bf z}_{n}^{j,\dag})$ and $\delta^\star$ by using a bisection
method\;

Calculate $D$ new vertices ${\bf{z}}_n^{j,d} $ by  \eqref{eqproj1} and update ${\mathbb V}_{j+1}$ by \eqref{eqVall}\;

{\bf   Bounding:} \\
Update the lower bound with the feasible inner point: ${\cal L}_n^{{\rm{Lb}}} \leftarrow \max \left\{ {{{\cal L}_n}\left( {{{\bf{a}}_n},{\rm{Proj}}\left( {{\bf{z}}_n^j} \right)} \right),{\cal L}_n^{{\rm{Lb}}}} \right\}$\;

Update  the upper bound by ${\bf{z}}_n^ {j,\dag} $, i.e., ${\cal L}_n^{{\rm{Ub}}} = {{\cal L}_n}\left( {{{\bf{a}}_n},{\bf{z}}_n^{j,\dag}} \right)$\;

\For {$\forall {\bf z}_n \in {\mathbb V}_{i+1}$} {\lIf{${{\cal L}_n}\left( {{{\bf{a}}_n},{\bf{z}}_n} \right)\le {\cal L}_n^{\rm Lb}$}{Delete ${\bf z}_n$ from ${\mathbb V}_{i+1}$ for memory and complexity reduction, i.e., ${{\mathbb V}_{j+1}} = {{\mathbb V}_{j+1}}\backslash \left\{ {\bf{z}}_n \right\}$ }}

$j=j+1$\;
}
}
\end{algorithm}

Finally, we summarize the above steps in Algorithm \ref{algorithmMO}. The MODP algorithm can be applied to find the optimal solution to problem \ref{P2}.

\subsection{Complexity Analysis}
We analyze the complexity of the proposed MODP algorithm, which comprises an outer DP framework and an inner MO solver.
Specifically, the overall time complexity is ${C_{{\rm{MODP}}}} = {\cal O}\left( {{N_{\rm{S}}}{N_{{\rm{ST}}}}{N_{\rm{A}}}{C_{{\rm{MO}}}}} \right)$
where ${{N_{\rm{S}}}}=N$ denotes the number of stages (time frames),
${{N_{{\rm{ST}}}}}$ is the number of DP state–transition evaluations per stage
${N_{\rm  A}}$ is the average number of admissible service-assignment actions for a given transition from ${{\mathbb S}_{n - 1}}$ to ${{\mathbb S}_{n}}$
and ${C_{{\rm{MO}}}}$ is the complexity of MO algorithm solving  \ref{P3a}.
Based on the state definition in \eqref{eqstate}, the state–transition count per stage is calculated by
\begin{align}
{N_{{\rm{ST}}}} = \sum\limits_{i = 0}^{\min \left( {K,NM'} \right)} {\left( {\left( {\begin{array}{*{20}{c}}
K\\
i
\end{array}} \right)\sum\limits_{j = 0}^{\min \left( {i,M'} \right)} {\left( {\begin{array}{*{20}{c}}
i\\
j
\end{array}} \right)} } \right)},
 \end{align}
where $M' = M{A_{\max }}$. Next, we analyze the complexity of the MO algorithm, which is dominated by two steps:
(i) maximum vertex determination, whose complexity scales with the variable dimension $D$,
and (ii) the projection step, which uses bisection over $\delta\in[0,1]$ with tolerance
$\varepsilon_b$ and therefore requires $O(\log_2(1/\varepsilon_b))$ iterations.
Consequently, we have ${C_{{\rm{MO}}}} = {\cal O}\left( {{D}{{\log }_2}\left( {\frac{{1}}{{{\varepsilon _b}}}} \right)} \right)$.

Although its complexity is lower than an exhaustive search, it still incurs high complexity. Nevertheless, this method serves as the optimal benchmark for other suboptimal algorithms.
Moreover, we note that the proposed MODP algorithm is applicable to other services whose VoS grows with increased resource allocation.

\section{Sub-Optimal Solution: VoS-Prioritized SCA}\label{Sec6}

In this section, we develop a suboptimal low-complexity VoS-prioritized SCA algorithm. It solves Problem \ref{P2} in two steps: the integer MDMA assignment is determined by a VoS-prioritization method, and the non-convex power allocation problem is tackled via SCA.

\subsection{VoS-prioritized MDMA Assignment}

\begin{algorithm}
\caption{VoS-prioritized MDMA Assignment}\label{algorithm3}
 \renewcommand{\baselinestretch}{0.8}
  {
Preliminary Distance-based Power Allocation:
\begin{align}&{p_{qmn}}= {\rm FixedPower}({\bf a},m,n) \nonumber\\
&= \left\{{\begin{array}{*{20}{l}}
{\frac{{d_q{P_{{\rm{max}}}}}}{{\sum\nolimits_{m = 1}^M\sum\nolimits_{j = 1}^{\left| {{{\mathbb K}_{\rm{C}}}} \right|\! +\! \left| {{{\mathbb K}_{\rm{P}}}} \right|} {d_j{a_{jmn}}} }},\;{\rm{if}}\;{a_{qmn}}\! = \! 1}\\
{0,\;{\rm{otherwise}}}
\end{array}} \right.\label{eqfixed}
 \end{align}

Initialize the unmatched user set by ${\mathbb K}^{\rm unU}={\mathbb K}$\;
Initialize the matched user set on RB $(m,n)$ as ${\mathbb K}_{mn}^{\rm RB}=\emptyset \; {\rm for}\;  \forall (m,n)$\;
Denote  $V_{kmn}^{\rm{X}} = \sum\nolimits_{i = 1}^{\left| {{{\mathbb Q}_{\rm{X}}}} \right|} {w_k^{{\rm{X}},i}} \log V_{kmn}^{{\rm{X}},i}$\;

\For{$A=1$ \KwTo $A_{\rm max}$}
     {      $\!$Initialize  the available RB set of user $j$ as $\!{\mathbb A}_j\!\!=\!\!\left\{\! {\left(\! {1\!,\!1}\! \right),\cdots,\left(\! {M\!,\!N} \!\right)} \!\right\}$  for $ 1\le j\le K$ when we consider each RB supports at most $A$ services \;
     $\!$\While{$\left| {{{\mathbb K}^{\rm unU}}} \right|>{\rm max}\left\{(K-AMN),0\right\}$}
      {$\!\!\!\!${Randomly select one service index $k$ from ${\mathbb K}^{\rm unU}$\;
        \For {$(m,n)=(1,1)$ \KwTo $(M,N)$}
            {\If{$\left| {{\mathbb K}_{mn}^{{\rm{RB}}}} \right| < A$ }
                { ${  {\widetilde{\mathbb K}}}_{mn}^{{\rm{RB}}} = {\mathbb K}_{mn}^{{\rm{RB}}} \cup \left\{ k \right\}$,
                ${{{ {\widetilde{\mathbb K}}}_{mn}^{{\rm{unU}}}}}={{{\mathbb K}^{{\rm{unU}}}}}\setminus\left\{ k \right\}$\;
                 Set ${a_{qmn}} = 1$ for $\forall q \in  {\widetilde{\mathbb K}}_{mn}^{{\rm{RB}}}$, otherwise ${a_{qmn}}=0$\;
                 Calculate  ${p_{qmn}}= {\rm FixedPower}({\bf a},m,n)$   and ${  V_{mn}} =\sum\nolimits_{q \in {\widetilde{\mathbb K}}_{mn}^{{\rm{RB}}}} {V_{qmn}^{\rm X}} $\;
                }
            \Else{ Set ${a_{qmn}} = 1$ for $\forall q \in {\widetilde{\mathbb K}}_{mn}^{{\rm{RB}}}$, otherwise ${a_{qmn}}=0$\;
                 Calculate  ${p_{qmn}}= {\rm FixedPower}({\bf a},m,n)$   and    ${{\widetilde V}_{mn} ^\star} = \sum\nolimits_{q \in {\widetilde{\mathbb K}}_{mn}^{{\rm{RB}}}} {V_{qmn}^{\rm X}} $ \;

             \For {$\forall j \in {{\mathbb K}_{mn}^{{\rm{RB}}}} $}

                  {$\tilde {\mathbb K}_{jmn}^{{\rm{RB}}}\! \!=\!\! {\mathbb K}_{mn}^{{\rm{RB}}} \!\cup\! \left\{ \!k\! \right\}\!\setminus\!\left\{\! j\! \right\}$,
                   ${{{\widetilde{\mathbb K}}_{jmn}^{{\rm{unU}}}}}\!\!=\!\!{{{\mathbb K}^{{\rm{unU}}}}}\!\cup\! \left\{ \!j\! \right\}\!\setminus\!\left\{ \!k \!\right\}$\;
                      Set ${a_{qmn}} = 1$ for $\forall q \in {{ \widetilde{\mathbb K}}_{jmn}^{{\rm{RB}}}}$, otherwise ${a_{qmn}}=0$\;
                 Calculate  ${p_{qmn}}= {\rm FixedPower}({\bf a},m,n)$   and     ${V_{mn}^j} = \sum\nolimits_{q \in {\widetilde {\mathbb K}}_{jmn}^{{\rm{RB}}}} {V_{qmn}^{\rm X}} $ \;
                      }
                   $V_{mn}^{{j^\star}} = \mathop {\arg \max }\nolimits_{\forall j \in {{  {\mathbb K}}_{mn}^{{\rm{RB}}}} } V_{mn}^j$\;
                             \If{${{\widetilde V}_{mn}^\star} > V_{mn}^{{j^\star}}$}
                                 { Set ${{\widetilde V}_{mn}} =  - \inf $,  $ {\widetilde {\mathbb K}}_{mn}^{{\rm{RB}}} = {\mathbb K}_{mn}^{{\rm{RB}}} $, ${{{\widetilde{\mathbb K}}_{mn}^{{\rm{unU}}}}} = {{{{\mathbb K}}^{{\rm{unU}}}}}$, and ${\mathbb A}_k={\mathbb A}_k\setminus\left\{(m,n)\right\}$ to avoid accommodate $k$ on RB $(m,n)$\;}
                             \Else{Set $\!\!{{  V}_{mn}}\! = \!V_{mn}^{{j^\star}}$, ${\widetilde {\mathbb K}}_{mn}^{{\rm{RB}}} \!= \!{\mathbb K}_{j^\star mn}^{{\rm{RB}}} $, ${{{ {\widetilde{\mathbb K}}}_{mn}^{{\rm{unU}}}}} \!= \!{{{{\mathbb K}}_{j^\star mn}^{{\rm{unU}}}}}$,  and ${\mathbb A}_{j^\star}={\mathbb A}_{j^\star}\setminus\left\{(m,n)\right\}$ to avoid accommodate $j^\star$ on RB $(m,n)$}

                }
            }
      $\left( {{m^\star},{n^\star}} \right) = \mathop {\arg \max }\nolimits_{\left( {m,n} \right)} {{V}_{mn}}$\
      ${ {\mathbb K}}_{{m^\star}{n^\star}}^{{\rm{RB}}}  = {\widetilde {\mathbb K}}_{{m^\star}{n^\star}}^{{\rm{RB}}}, {\mathbb K}^{\rm unU}={\widetilde{\mathbb K}}^{\rm unU}_{m^\star n^\star}$
         }
}
}}
\end{algorithm}

In this part, we develop a heuristic algorithm to find the sub-optimal MDMA assignment on each RB by using VoS prioritization with fixed power allocation.
The design is motivated by the practical constraint that the number of services typically exceeds the available RBs, necessitating NOMA to support multiple services per RB. Moreover, assigning multiple services to a single RB introduces resource sharing, which increases interference and may degrade performance.
Therefore, we propose a heuristic VoS-prioritized MDMA assignment algorithm that iteratively allocates services to RBs up to the maximum multiplexing limit. Consequently, we sequentially process each unassigned service, evaluate its achievable VoS across all RBs, and assign it to the RB that yields the highest VoS. The evaluation process incorporates the following key aspects:
\begin{itemize}
\item Since evaluating VoSs under a specific MDMA assignment requires a predetermined power allocation, we apply a fixed distance-based power allocation scheme to evaluate the achievable VoSs.

\item If the selected RB has not reached its multiplexing limit, the VoS of the selected RB is evaluated by incorporating the new service into the existing set.

\item
If the RB has reached full capacity, the algorithm evaluates whether swapping any one of the currently allocated services with the new candidate would improve the total VoS of the RB. The swap is executed only if it results in a VoS improvement. In such cases, the removed service is excluded from future allocation to this RB. If no swap occurs during this evaluation, the candidate service is excluded from this RB in subsequent iterations.

\end{itemize}
This iterative process gradually increases service assignments per RB while ensuring all services are allocated to suitable RBs. The complete algorithm, detailed in Algorithm \ref{algorithm3}, provides a computationally efficient approach to VoS maximization under multiplexing constraints.

\subsection{SCA-based Power Allocation}\label{secPower}
Given the MDMA assignment in problem \ref{P2}, the remaining problem is the power allocation, which remains non-convex. Therefore, we propose a low-complexity sub-optimal algorithm to solve it efficiently.

To begin with, we re-express the non-convex constraints \eqref{eqp2c1} and \eqref{eqp2c3} as the following equivalent forms:
\begin{align}
\frac{1}{4}\left( {{\cal Q}_{qkmn}^A\left( {{\bf{z}},{\bf{p}}} \right) - {\cal Q}_{qkmn}^B\left( {{\bf{z}},{\bf{p}}} \right)} \right) \le {p_{kmn}}\chi _{qkmn}^{\rm{C}},\nonumber\\
{\rm{if}}\;{a_{qmn}} = 1,1 \le q \le k \le \left| {{{\mathbb K}_{\rm{C}}}} \right|, \label{eqsca1}\\
\frac{1}{4}\left( {{\cal Q}_{kmn}^C\left( {{\bf{z}},{\bf{p}}} \right) - {\cal Q}_{kmn}^D\left( {{\bf{z}},{\bf{p}}} \right)} \right) \le BL{p_{kmn}}{\lambda _{kmn}},\nonumber\\
k \in {{\mathbb K}_{\rm{S}}},\label{eqsc2}
\end{align}
where  
\begin{align}
{\cal Q}_{qkmn}^A\left( {{\bf{z}},{\bf{p}}} \right) &= {( {z_{kmn}^{\rm{C}} +  {\sum\limits_{j = 1}^{k - 1} {{a_{jmn}}{p_{jmn}}\chi _{qjmn}^{\rm{C}}}  + {\sigma _q}} } )^2}, \\
{\cal Q}_{qkmn}^B\left( {{\bf{z}},{\bf{p}}} \right)& = {( {z_{kmn}^{\rm{C}} - {\sum\limits_{j = 1}^{k - 1} {{a_{jmn}}{p_{jmn}}\chi _{qjmn}^{\rm{C}}}  - {\sigma _q}} } )^2},
\end{align}
\begin{align}
{\cal Q}_{kmn}^C\left( {{\bf{z}},{\bf{p}}} \right)& = {( {z_{kmn}^{\rm{S}} +\!\! {\sum\limits_{k' \in {{\mathbb K}_{\rm{C}}} \cup {{\mathbb K}_{\rm{P}}}}  \!\!{a_{k'mn}}{p_{k'mn}}\chi _{kk'mn}^{\rm{S}} + {\sigma _k}} } )^2},
\end{align}
\begin{align}
{\cal Q}_{kmn}^D\left( {{\bf{z}},{\bf{p}}} \right)& = {( {z_{kmn}^{\rm{S}} - \!\! {\sum\limits_{k' \in {{\mathbb K}_{\rm{C}}} \cup {{\mathbb K}_{\rm{P}}}}  \!\!{a_{k'mn}}{p_{k'mn}}\chi _{kk'mn}^{\rm{S}} - {\sigma _k}} } )^2}.
\end{align}

Obviously, ${\cal Q}_{qkmn}^A\left( {{\bf{z}},{\bf{p}}} \right)$,    ${\cal Q}_{qkmn}^B\left( {{\bf{z}},{\bf{p}}} \right)$, ${\cal Q}_{kmn}^C\left( {{\bf{z}},{\bf{p}}} \right)$, and ${\cal Q}_{kmn}^D\left( {{\bf{z}},{\bf{p}}} \right)$ are convex functions.
Thus,  this problem belongs to the Difference of Convex (DC) programming family, where all non-convex components can be expressed as differences between convex/concave functions while maintaining convexity in the remaining structure.
The problem is therefore solved using the SCA algorithm. This method iteratively approximates non-convex terms through first-order Taylor expansions, transforming the original problem into a sequence of convex subproblems. Since each subproblem is convex, it can be solved to optimality. Furthermore, as successive approximations are constructed to build upon previous optimal solutions while ensuring monotonic improvement in the objective value, the algorithm's convergence is guaranteed \cite{chen2019resource}.

Upon denoting the optimized solutions of ${\bf p} $ and ${\bf z} $ in  the $i$-th iteration  by ${\bf p}_{\left[i\right]}$ and ${\bf z}_{\left[i\right]} $, respectively,  we have
\begin{align}
{\cal Q}_{qkmn}^B\left( {{\bf{z}},{\bf{p}}} \right)
\ge& {\cal Q}_{qkmn}^B\left( {{{\bf{z}}_{\left[i\right]}},{{\bf{p}}_{\left[i\right]}}} \right) \!+\! \nabla {\cal Q}_{qkmn}^B{\left( {{{\bf{z}}_{\left[i\right]}},{{\bf{p}}_{\left[i\right]}}} \right)^{\!\rm{T}}}\left[\!\! {\begin{array}{*{20}{c}}
{\bf{z}}\\
{\bf{p}}
\end{array}} \!\!\right] \nonumber\\
 \buildrel \Delta \over = &{\cal Q}_{qkmn}^{B,{\rm{lb}}}\left( {{\bf{z}},{\bf{p}}\left| {{{\bf{z}}_{\left[i\right]}},{{\bf{p}}_{\left[i\right]}}} \right.} \right),\\
 {\cal Q}_{kmn}^D\left( {{\bf{z}},{\bf{p}}} \right)
 \ge& {\cal Q}_{kmn}^D\left( {{{\bf{z}}_{\left[i\right]}},{{\bf{p}}_{\left[i\right]}}} \right)\!+\! \nabla {\cal Q}_{kmn}^D{\left( {{{\bf{z}}_i},{{\bf{p}}_i}} \right)^{\!\rm{T}}}\left[\!\! {\begin{array}{*{20}{c}}
{\bf{z}}\\
{\bf{p}}
\end{array}}\!\! \right] \nonumber \\
 \buildrel \Delta \over =& {\cal Q}_{kmn}^{D,{\rm{lb}}}\left( {{\bf{z}},{\bf{p}}\left| {{{\bf{z}}_{\left[i\right]}},{{\bf{p}}_{\left[i\right]}}} \right.} \right),
\end{align}
where $\nabla {\cal Q}_{qkmn}^B{\left( {{{\bf{z}}_{\left[i\right]}},{{\bf{p}}_{\left[i\right]}}} \right)^{\!\rm{T}}}$ and $\nabla {\cal Q}_{kmn}^D{\left( {{{\bf{z}}_i},{{\bf{p}}_i}} \right)^{\!\rm{T}}}$ are the gradients of functions $
{\cal Q}_{qkmn}^B\left( {{\bf{z}},{\bf{p}}} \right)$ and  $ {\cal Q}_{kmn}^D\left( {{\bf{z}},{\bf{p}}} \right)$ at the point  $\left({\bf p}_{\left[i\right]}, {\bf z}_{\left[i\right]} \right)$, respectively.

Then, the power optimization of \ref{P2} can be reformulated as the following form  in the $(i+1)$-th iteration of the SCA, i.e.,
\vspace{-0.5cm}
\begin{subequations}
\begin{align}
\!\!\mathop {\max }\limits_{{\bf{p}},{\bf{z}}} \;& {\cal L}\left({\bf a}, {\bf z}\right) \tag{\bf P4} \label{P4}\\
{\rm{s}}{\rm{.t}}{\rm{.}}\;
& \frac {1}{4}{\left( {{\cal Q}_{qkmn}^A\left( {{\bf{z}},{\bf{p}}} \right) - {\cal Q}_{qkmn}^{B,{\rm lb}}\left( {{\bf{z}},{\bf{p}}} \left| {{{\bf{z}}_{\left[i\right]}},{{\bf{p}}_{\left[i\right]}}} \right.\right)} \right)} \nonumber\\
&\le {p_{kmn}}\chi _{qkmn}^{\rm{C}},{\rm{if}}\;{a_{qmn}} = 1,1 \le q \le k \le \left| {{{\mathbb K}_{\rm{C}}}} \right|, \label{eqsca12}\\
&\frac{1}{4}\left( {{\cal Q}_{kmn}^C\left( {{\bf{z}},{\bf{p}}} \right) - {\cal Q}_{kmn}^{D,{\rm lb}}\left( {{\bf{z}},{\bf{p}}} \left| {{{\bf{z}}_{\left[i\right]}},{{\bf{p}}_{\left[i\right]}}} \right.\right)} \right) \nonumber\\
&\qquad\qquad\qquad\qquad\quad\le BL{p_{kmn}}{\lambda _{kmn}},k \in {{\mathbb K}_{\rm{S}}},\label{eqsc22}\\
&{\bf{z}}_n^{{\rm{min}}}\le{\bf{z}}_n, \forall n, \label{eqsc31} \\
&\eqref{PFCA}-\eqref{PFCE}, \eqref{eqp2c2},{\rm and}\; \eqref{eqp2c4}.
\end{align}
\end{subequations}
Here, \eqref{eqsc31}  is introduced to ensure the concavity of the objective function, as stated in Theorem \ref{lemma4}.
Besides, given the proportional VoS framework adopted in this study and the definition of  ${\cal V}\left( {Q,\tilde Q, \alpha, \beta } \right)$  in \eqref{equationQos2a} and \eqref{equationQos2b}, the objective function value will be negative infinity if this condition is not satisfied.
Hence, the inclusion of this new constraint does not reduce the value of the objective function.

Moreover, combined with the results of Theorem \ref{lemma4}, we conclude that \ref{P4} is a convex optimization problem that, in principle, can be solved efficiently using standard convex optimization methods such as interior-point algorithms or the Matlab toolbox CVX.
However, due to the complex expression of the concave objective function in problem \ref{P4} and its incompatibility with CVX’s convexity requirements, we can employ a linear interpolation approach to approximate the objective function using a set of linear functions.
This approximation enables the effective application of CVX to solve the problem.
After this, we can further enhance overall performance by refining the MDMA assignment with a swap operation to improve the total VoS across all RBs.
The complete procedure is summarized in Algorithm \ref{algorithmSCA}.

\subsection{Complexity Analysis}

We now analyze the computational complexity of the proposed VoS-prioritized SCA algorithm. The overall complexity consists of two dominant components:
the VoS-prioritized MDMA assignment complexity, denoted by $C_{\mathrm{VoS}}$, and the SCA optimization complexity, denoted by  $C_{\mathrm{SCA}}$.
Accordingly, the total complexity is $C_{\mathrm{VoS-SCA}} = C_{\mathrm{VoS}} + C_{\mathrm{SCA}}$.
Specifically, the assignment complexity $C_{\mathrm{VoS}}$ is governed by three key parameters: the number of services $K$, the number of RBs $M \times N$, and the maximum concurrency limit $A_{\max}$.
In the worst-case scenario, when an RB reaches its capacity limit, the algorithm performs up to $A_{\max}$ swap-out evaluations. Therefore,  the overall assignment complexity can be given by $ C_{\mathrm{VoS}} = \mathcal{O}\left(KMNA_{\max}\right)$.
Next, the SCA complexity $C_{\mathrm{SCA}}$ is primarily determined by three factors: the number of iterations $N_{\mathrm{it}}$, the dimension of the optimization variables $N_V$, and the number of constraints $N_C$. From \cite{chen2024otfs}, this relationship can be expressed by
$C_{\mathrm{SCA}} = \mathcal{O}\left(N_{\mathrm{it}} N_{\rm C}^{1.5}N_{\rm V}^2\right)$.

\section{Simulations}\label{Sec7}

\begin{algorithm}[t]{
\caption{Joint Swap and SCA for \ref{P2}}\label{algorithmSCA}
\renewcommand{\baselinestretch}{0.92}
{
\While {No further new swap can increase the total VoS and the maximum number of iterations has not been reached}
    {
    $\!\!\!\!\!${\bf Enhance the MDMA assignment by Swap}\\

 With the obtained $\bf a$,  perform a swap operation, i.e., randomly change $a_{kmn}$ from 1 to 0 and simultaneously change $a_{km'n'}$ from 0 to 1 for $(m,n) \neq (m',n')$, subject to constraint \eqref{PFCC}\;

   $\!\!\!\!\!${\bf Given MDMA assignment, optimize the power allocation by SCA }\\

    \While{ $\left|  {\cal L}\left({\bf a}, {\bf z}_{\left[i\right]}\right)- {\cal L}\left({\bf a}, {\bf z}_{\left[i-1\right]}\right) \right| \le \bar\varepsilon $ }{
    Given ${\bf z}_{\left[i\right]}$ and ${\bf p}_{\left[i\right]}$, calculate ${\bf z}_{\left[i+1\right]}$ and ${\bf p}_{\left[i+1\right]}$ by solving  problem \ref{P4}\;
    $i \leftarrow i+1$\;
    }

  \lIf {the new swap yields a better objective function value for Problem \ref{P2}}
    {
        Update the current $\mathbf{a}$ accordingly
    }
    \lElse
    {
        Retain the current $\mathbf{a}$
    }
    }


} }
\vspace{-0.1cm}
\end{algorithm}

The system is operated on a carrier frequency of $f_c=5.9$ GHz, and the subcarrier bandwidth $\Delta_f=156.25$ KHz in accordance with the IEEE 802.11p \cite{nguyen2017delay}. The OFDM symbol duration, including the cyclic prefix, is set to $T=8$ us. The numbers of sub-carriers per RB  and symbols per RB are set to  $B=8$ and $L=8$, respectively.
Then, we assume that the BS is located at the origin of a two-dimensional coordinate plane.
The angle  $\theta_k$  is uniformly distributed in $\left[-\pi/3,\pi/3\right]$, and the distance for communication/sensing and positioning users are uniformly distributed
in $\left[30{\rm m}, 1000{\rm m}\right]$ and $\left[30{\rm m}, 200{\rm m}\right]$, respectively.
We assume that channel ${\bf h}_{kmn}$ follows Rician fading, where the
 path-loss is modeled as  $74.2+16.8\lg(d_k/{\rm 1m})$, and the Rician factor is set to $1$.
The maximum powers at sensing users are set to $-5$ dBm, and the noise powers at the BS or users are set to -114 dBm.
The RCS for $k\in{{\mathbb K}_{\rm{P}}} \cup {{\mathbb K}_{\rm{S}}}$ is set to $\delta _k^{{\rm{RCS}}}=1$.
The priority weights $w_k^{{\rm{X}},i}$  are randomly generated from a uniform distribution over the interval [0, 1].
For \mbox{Type-$\rm C$} services, the target KPIs are set to $\tilde Q_{k}^{{\rm C},1}=4$ bits/s/Hz and $\tilde Q_{k}^{{\rm C},2}=LT$.
For \mbox{Type-$\rm P$} services,  the target KPIs are set to $\tilde Q_{k}^{{\rm C},1}=  \frac{1}{20}{I_{kmn}^\theta }  $, $\tilde Q_{k}^{{\rm C},2}=\frac{1}{20}{I_{kmn}^d }  $, $\tilde Q_{k}^{{\rm C},3}=\frac{1}{20}{I_{kmn}^v }$, and $\tilde Q_{k}^{{\rm C},4}=LT$, respectively.
For \mbox{Type-$\rm S$} services, we set $r_k=30$ m and ${{\cal P}_{k}^{{\rm{FA}}}}=0.3$. Then, the target KPIs are set to $\tilde Q_{k}^{{\rm S},1}=0.8$ and $\tilde Q_{k}^{{\rm S},2}=LT$. Moreover, the elasticity parameters $\alpha _k^{{\rm{X}},i}$ and $\beta _k^{{\rm{X}},i}$ for $\forall k, \forall i$, and $ \forall {\rm X}$ are uniformly distributed in $\left[0,\alpha\right]$ and $\left[0,\beta\right]$, respectively.
Additionally, we use the maximum ratio transmission (MRT) beamforming vectors in all simulations.
Finally, we consider the following baselines for performance comparison to validate the proposed designs:
1) VoS-Fixed: VoS prioritization is applied to optimize MDMA design, followed by fixed power allocation mentioned in  \eqref{eqfixed};
2) Random-SCA: random MDMA assignment is implemented with the SCA proposed in Section~\ref{secPower}  for power allocation;
3) Random-Fixed: random MDMA assignment is implemented with fixed power allocation mentioned in  \eqref{eqfixed}.

\begin{figure}[t]
\vspace{-0.25cm}
   \centering
   \includegraphics[trim=20 1.7 30 18, clip,width=.42\textwidth]{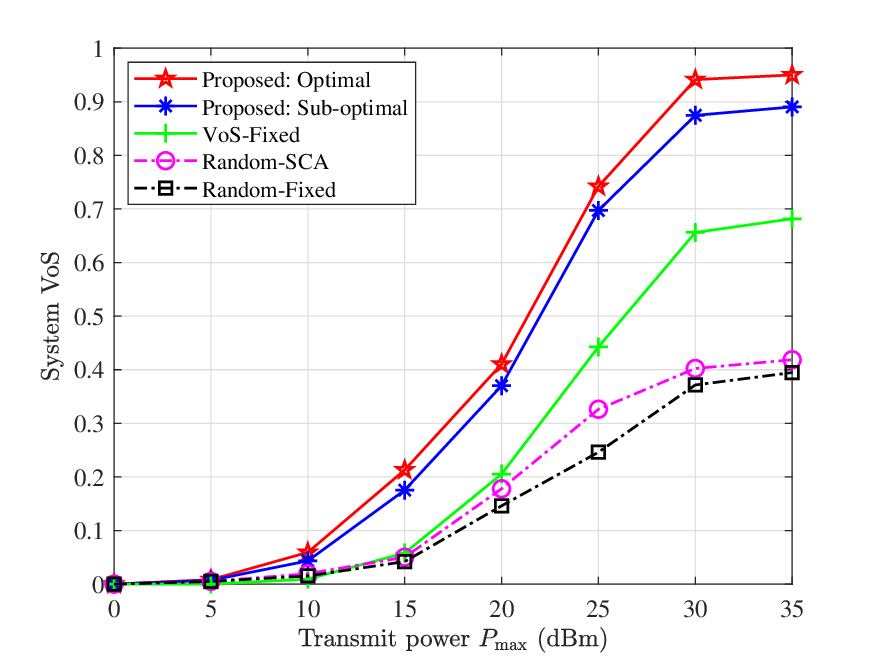}
  \caption{The impact of the maximum transmission power on the system VoS: $\left| {{{\mathbb K}_{\rm C}}} \right|=3$, $\left| {{{\mathbb K}_{\rm P}}} \right|=2$,   $\left| {{{\mathbb K}_{\rm S}}} \right|=1$, $M=1$, $N=3$, $L_{\rm tx}=4$, $A_{\rm max}=2$,  $\alpha=0.3$, and $\beta=0.3$.}\label{fig4Sim} 
 \end{figure}

\begin{figure}[t]
   \centering
   \includegraphics[trim=20 1.7 30 18, clip,width=.42\textwidth]{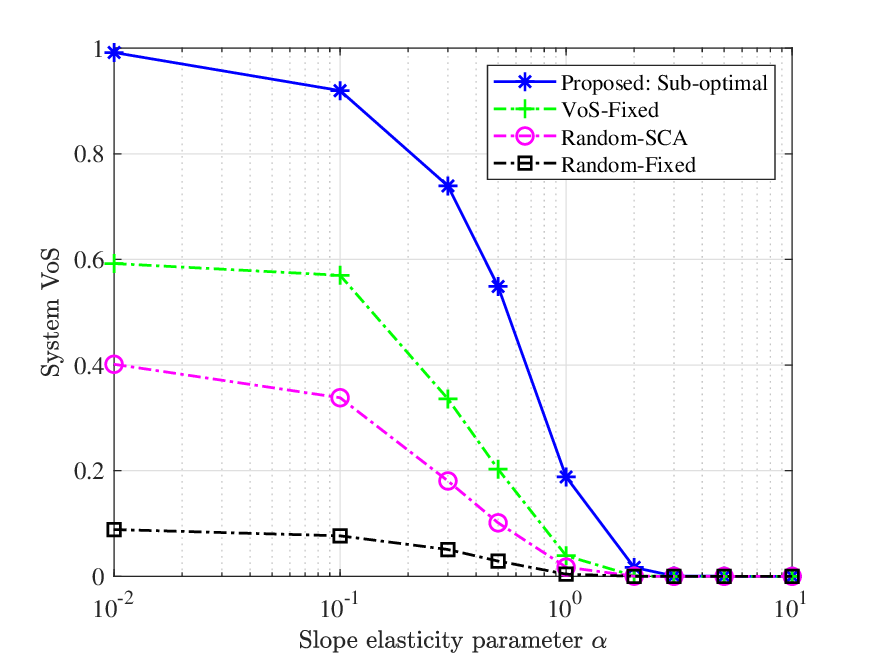}
 \caption{The impact of the slope elasticity parameter on the system VoS: $\left| {{{\mathbb K}_{\rm C}}} \right|=6$, $\left| {{{\mathbb K}_{\rm P}}} \right|=5$,   $\left| {{{\mathbb K}_{\rm S}}} \right|=4$, $M=2$, $N=3$, $L_{\rm tx}=4$, $A_{\rm max}=4$,  $P_{\rm max}= 30$ dBm, and $\beta=0.2$.}\label{fig5Sim} 
 \end{figure}

 \begin{figure}[t]
   \centering
   \includegraphics[trim=20 1.7 30 18, clip,width=.42\textwidth]{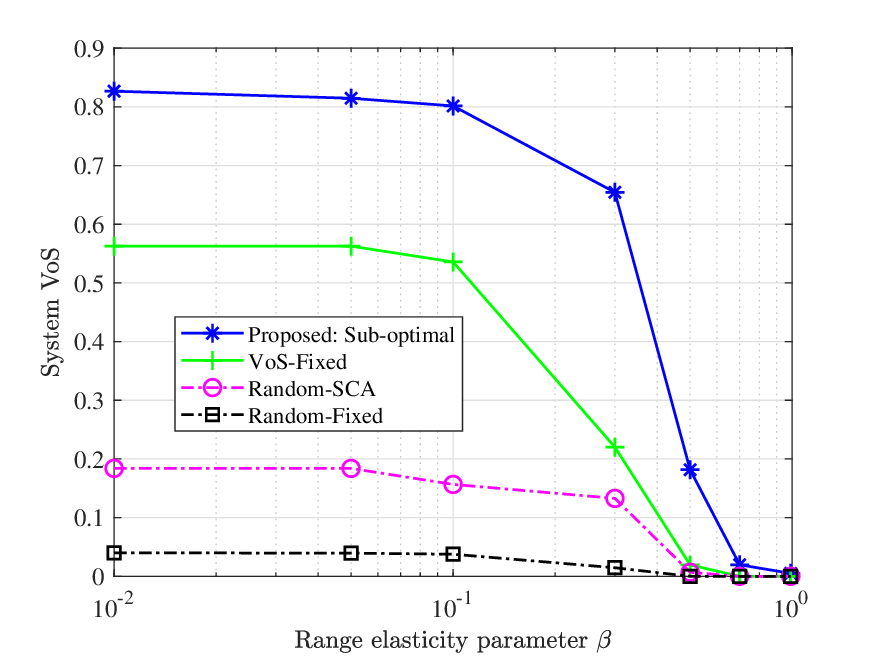}
   \caption{The impact of the range elasticity parameter on the system VoS: $\left| {{{\mathbb K}_{\rm C}}} \right|=6$, $\left| {{{\mathbb K}_{\rm P}}} \right|=5$,   $\left| {{{\mathbb K}_{\rm S}}} \right|=4$, $M=2$, $N=3$, $L_{\rm tx}=4$, $A_{\rm max}=4$,  $P_{\rm max}= 30$ dBm, and $\alpha=0.3$.}\label{fig6Sim} 
 \end{figure}

  \begin{figure}[t]
   \centering
   \includegraphics[trim=20 1.7 30 18, clip,width=.42\textwidth]{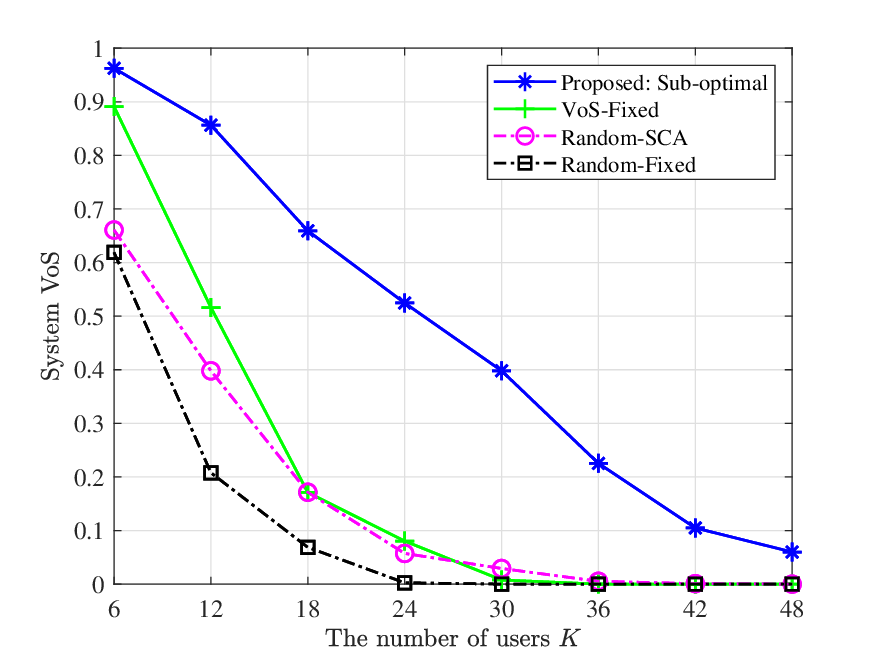}
\caption{The impact of the number of users on the system VoS: $\left| {{{\mathbb K}_{\rm C}}} \right|=\left| {{{\mathbb K}_{\rm P}}} \right|=\left| {{{\mathbb K}_{\rm S}}} \right|=K/3$, $M=3$, $N=3$, $L_{\rm tx}=4$, $A_{\rm max}=6$,  $P_{\rm max}= 30$ dBm, $\alpha=0.3$, and  $\beta=0.3$.}\label{fig7Sim} 
 \end{figure}

   \begin{figure}[t]
   \centering
   \includegraphics[trim=20 1.7 30 18, clip,width=.42\textwidth]{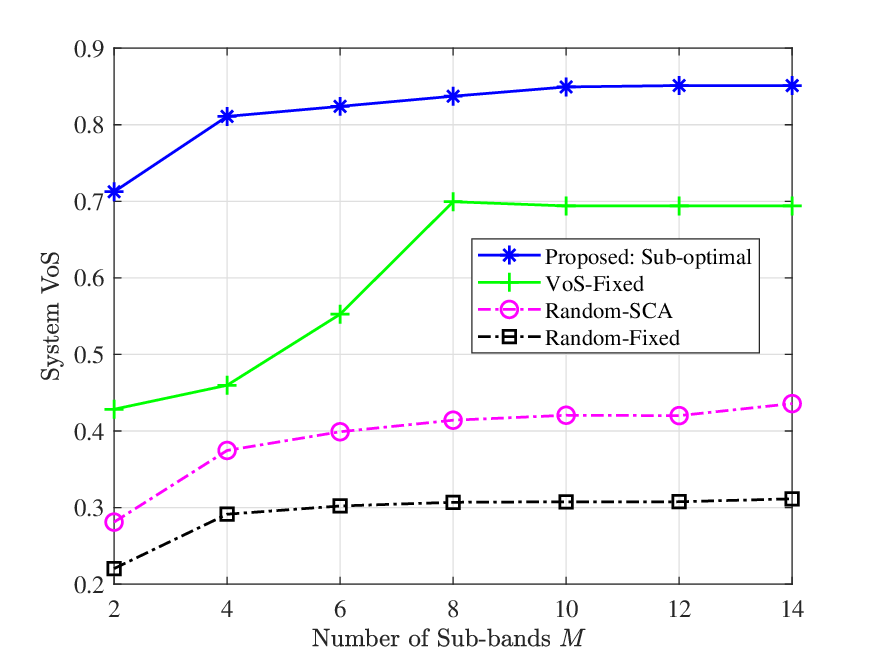}
  \caption{The impact of the number of sub-bands on the system VoS: $\left| {{{\mathbb K}_{\rm C}}} \right|=\left| {{{\mathbb K}_{\rm P}}} \right|=\left| {{{\mathbb K}_{\rm S}}} \right|=3$,   $N=2$, $L_{\rm tx}=4$, $A_{\rm max}=3$,  $P_{\rm max}= 30$ dBm, $\alpha=0.8$, and  $\beta=0.1$.}\label{fig8Sim} 
 \end{figure}
Fig. \ref{fig4Sim} examines the impact of transmission power on the proportional VoS of all users.
As the power increases, the VoS of all algorithms increases because higher power reduces the performance loss relative to the target KPI values.
Additionally, the proposed sub-optimal algorithm achieves performance close to that of the optimal algorithm while outperforming the other algorithms, thus demonstrating its effectiveness.

Fig. \ref{fig5Sim} investigates the effect of the slope elasticity parameter $\alpha$, which determines how performance loss in a given KPI impacts the final normalized value.
The performance of all algorithms decreases as $\alpha$ increases, since a larger $\alpha$ reduces elasticity, resulting in a steeper slope and greater degradation of the normalized value.
Moreover, the overall VoS gradually approaches zero due to the diminished elasticity.
Nevertheless, the proposed algorithm consistently outperforms the others, thus validating its effectiveness.

Fig. \ref{fig6Sim} investigates the effect of the range elasticity parameter $\beta$, which defines the meaningful range of the difference between the target and achieved KPI values.
Performance decreases as $\beta$ increases, since a larger $\beta$ reduces elasticity in this range, increasing the likelihood of performance degradation in terms of the normalized value.
Moreover, the overall VoS of all algorithms eventually approaches zero due to diminished elasticity.
The proposed algorithm consistently outperforms the others, further validating its effectiveness.

Fig. \ref{fig7Sim} investigates the effect of the number of users on the proportional VoS across all users.
The results show that the performance of all algorithms decreases as the number of users increases.
This decline is attributed to the nature of proportional VoS, where each additional user requests network resources, leading to increased interference and reduced performance for all users.
Ultimately, the VoS approaches zero when available radio resources are insufficient to support the growing number of users.
Notably, the proposed algorithm consistently outperforms the baseline scheme, further validating its effectiveness.

Fig. \ref{fig8Sim} investigates the effect of the number of sub-bands on the proportional VoS across all users.
As this number increases, performance improves due to the enhanced degree of freedom in the optimization process and additional spectrum resources involved.
However, performance eventually plateaus due to the limited power resources. Beyond a certain point, additional freedom does not lead to further performance gains.

\section{Conclusions}\label{Sec8}

This paper has proposed an elastic VoS-prioritized MDMA mechanism for IHSP systems.
Specifically, we have developed a comprehensive VoS metric that prioritizes the significance of completing a service amidst competing diverse service provisioning.
We have then leveraged the MDMA scheme to flexibly accommodate services across multi-dimensional RBs, considering user-specific interference tolerances and interference cancellation capabilities.
Subsequently, we have formulated a proportional VoS maximization problem by jointly optimizing MDMA design and power allocation.
Although the problem is non-convex, we have efficiently addressed it by introducing both an optimal MODP algorithm and a suboptimal VoS-SCA algorithm.
Finally, simulation results have confirmed the effectiveness of the proposed schemes, demonstrating that the modified VoS, which incorporates elastic parameters to account for user-specific performance tolerances across each KPI, serves as a robust and effective metric. Furthermore, the VoS-based elastic MDMA scheme has proved to be an efficient MA strategy for addressing the diverse, stringent, and competing demands within IHSP.

\appendix

\subsection{Proof of Theorem~\ref{theorem1}} \label{appendix1}

Upon assuming that the positioning users are widely separated in the surveillance region, the CRB matrix of multiple users is a block diagonal matrix, and the CRB for each user can be approximately and individually derived from the single target case of \eqref{eqCRLBsignalA} \cite{chen2024learning}. Then, from \eqref{eqCRLBsignalA}, with some algebraic transformations, we have
\begin{align}
\tilde y_{kmn}^{{\rm{P}},bl\ell}& = \frac{{\bar y_{mn}^{{\rm{P}},bl\ell}{{\left( {\bar s_{kmn}^{bl}} \right)}^*}}}{{{{\left| {\bar s_{kmn}^{bl}} \right|}^2} }} \nonumber\\
&  \approx  {\rho _{kmn}}{e^{{\rm{j}}{\phi _{kmn}}}}{e^{{\rm{j}}\pi \left[ {2\left( {{\nu_{km}}lT - b{\Delta _f}{\tau _{k}}} \right) - \ell \sin {\theta _{k}}} \right]}} + \tilde u_{kmn}^{{\rm{P}},bl\ell}, \nonumber\\
&  =  {\rho _{kmn}}{e^{{\rm{j}}\left[ {{\phi _{kmn}} + \ell \sin {{\bar \theta }_k} + l{{\bar \nu }_{km}} + b{{\bar \tau }_k}} \right]}} + \tilde u_{kmn}^{{\rm{P}},bl\ell },\label{eqCRB1}
\end{align}
where $\tilde u_{kmn}^{{\rm{P}},bl\ell} = \frac{{u_{mn}^{{\rm{P}},bl\ell}{{\left( {\bar s_{kmn}^{bl}} \right)}^*}}}{{{{\left| {\bar s_{kmn}^{bl}} \right|}^2} }}$ is approximated as a CSCG variable with mean zero and covariance $\frac{1}{{z _{kmn}^{\rm{P}}}}=\frac{{{\sigma _0}}}{{\sum\nolimits_{k' \in {{\mathbb K}_{\rm{C}}} \cup {{\mathbb K}_{\rm{P}}}} {{a_{k'mn}}{p_{k'mn}}\chi _{kk'mn}^{\rm{P}}} }}$. Besides, $\bar\theta_k=-\theta_k$, ${{\bar \tau }_k} =- 2\pi {\Delta _f}{\tau _k}$ and ${{\bar \nu }_{km}} = 2\pi T{\nu _{km}}$.

Then, upon denoting ${{{{\bm \eta}}}_{kmn}} = \left[ {{\bar \theta _k},{{\bar \tau }_k},{{\bar \nu }_{km}},{\rho _{kmn}},{\phi _{kmn}}} \right]$,  the   Fisher information matrix (FIM) for user $k$ within $(m,n)$-th RB is
\begin{align}
{{\bf{F}}_{kmn}} = 2z_{kmn}^{\rm{P}}\underbrace {\sum\limits_{\ell  = 0}^{{L_{{\rm{tx}}}} - 1} {\sum\limits_{b = 0}^{B - 1} {\sum\limits_{l = 0}^{L - 1} {{\bf{J}}_{kmn}^{bl\ell }} } } }_{{{\bf{J}}_{kmn}}}\in{\mathbb C}^{5\times5},\label{eqCRB2}
\end{align}
where the $(i,j)$-th element  of the matrix  ${{\bf{J}}_{kmn}^{bl\ell }}$ is given by
\begin{align}
&{\left[ {{\bf{J}}_{kmn}^{bl\ell }} \right]_{ij}} = \nonumber\\
&\frac{{\partial \Re \left( {\tilde y_{kmn}^{{\rm{P}},bl\ell }} \right)}}{{{\bm{\eta }}_{kmn}^i}}\frac{{\partial \Re \left( {\tilde y_{kmn}^{{\rm{P}},bl\ell }} \right)}}{{{\bm{\eta }}_{kmn}^j}} + \frac{{\partial \Im \left( {\tilde y_{kmn}^{{\rm{P}},bl\ell }} \right)}}{{{\bm{\eta }}_{kmn}^i}}\frac{{\partial \Im \left( {\tilde y_{kmn}^{{\rm{P}},bl\ell }} \right)}}{{{\bm{\eta }}_{kmn}^j}}.\label{eqCRB3}
\end{align}
Here, ${{\bm{\eta }}_{kmn}^i}$ is the $i$-th entry of  ${\bm{\eta }}_{kmn}$ for $1\le i\le 5$.
Besides, $\Re \left(  \cdot  \right)$  and $\Im \left(  \cdot  \right)$ denote the real and imaginary components of the input number, respectively.

With \eqref{eqCRB1},
${\left[ {{\bf{J}}_{kmn}^{bl\ell }} \right]_{ij}}$ defined in \eqref{eqCRB3} can be calculated  as \eqref{eqJkmn}.
Then, we know
\begin{align}
\!{\rm{E}}\left({\left| {{\bm{\eta }}_{kmn}^i \!-\! {\bm{\hat \eta }}_{kmn}^i} \right|^2}\right) \ge {\left[ {{\bf{F}}_{kmn}} \right]_{ii}^{ - 1}}\! = \! \frac{1}{{2z_{kmn}^{\rm{P}}}}{\left[ {{\bf{J}}_{kmn}} \right]_{ii}^{ - 1}},
\end{align}
where ${\left[ {{\bf{F}}_{kmn}} \right]_{ii}^{ - 1}} $ and ${\left[ {{\bf{J}}_{kmn}} \right]_{ii}^{ - 1}}$  for $1\le i\le5$ denote the $(i,i)$-th elements of
 ${\left[ {{\bf{F}}_{kmn}} \right]^{ - 1}} $ and ${\left[ {{\bf{J}}_{kmn}} \right]^{ - 1}}$, respectively.

Next, since  $\bar\theta_k=-\theta_k$, ${{\bar \tau }_k} =- 2\pi {\Delta _f}{\tau _k}$ with ${\tau _k} = \frac{{2{d_k}}}{{{c_o}}}$, and ${{\bar \nu }_{km}} = 2\pi T{\nu _{km}}$ with ${\nu _{km}} = \frac{{2{v_k}{f_m}}}{{{c_o}}}$, it follows \eqref{eq12} and  Theorem~\ref{theorem1} is proved.

\subsection{Proof of Lemma~\ref{lemmaA}}\label{appendix2}
First,  we know that $\log(1+x)$ is concave and
nondecreasing in $x>0$ and $\frac{1}{x}$ is convex and nonincreasing in $x>0$.
Based on these properties, the corresponding results for the Type-C and Type-P services can be derived directly.

Next, we provide the proof for the results of Type-S services.
From \eqref{eq21}, the first derivative of ${Q_{kmn}^{{\rm{S}},1}}$ with respect to ${z_{kmn}^{\rm{S}}}$ can be calculated as:
\begin{align}
\frac{{{\rm{d}}Q_{kmn}^{{\rm{S}},1}}}{{{\rm{d}}z_{kmn}^{\rm{S}}}} = \frac{{{W_k}}}{{2{{\left( {z_{kmn}^{\rm{S}} + 1} \right)}^2}}}{\cal G}\left( {z_{kmn}^{\rm{S}}} \right) > 0,
  \end{align}
where ${{W_k}} = F_{{\varpi _k}}^{ - 1}\left( 1-{{\cal P}_{k}^{{\rm{FA}}}} \right)$ and ${\cal G}\left( {z_{kmn}^{\rm{S}}} \right)= {e^{ - \frac{1}{2}\left( {\frac{{W_k}}{{z _{kmn}^{\rm{S}} + 1}}} \right)}}$. Thus, it is a non-decreasing function of ${z_{kmn}^{\rm{S}}}$. Besides, the second derivative of ${Q_{kmn}^{{\rm{S}},1}}$  with respect to ${z_{kmn}^{\rm{S}}}$ is
\begin{align}
\frac{{{{\rm{d}}^2}Q_{kmn}^{{\rm{S}},1}}}{{{\rm{d}}{{\left( {z_{kmn}^{\rm{S}}} \right)}^2}}} =  - \frac{{{W_k}\left( {4\left( {z_{kmn}^{\rm{S}} + 1} \right) - {W_k}} \right)}}{{4{{\left( {z_{kmn}^{\rm{S}} + 1} \right)}^4}}}{\cal G}\left( {z_{kmn}^{\rm{S}}} \right).
  \end{align}
  When $\frac{1}{e^2} \leq {\cal P}_k^{\rm{FA}}$, it holds that  $4\left( {z_{kmn}^{\rm{S}} + 1} \right) - W_k \ge 4-W_k= 4 + 2\ln {\cal P}_k^{{\rm{FA}}} \ge0 $ for ${z_{kmn}^{\rm{S}}}\ge0$.
  Consequently, we have $
\frac{{{{\rm{d}}^2}Q_{kmn}^{{\rm{S}},1}}}{{{\rm{d}}{{\left( {z_{kmn}^{\rm{S}}} \right)}^2}}}\le0$  and the corresponding concavity is proved.
 Therefore, Lemma~\ref{lemmaA} is proved.

\subsection{Proof of Lemma~\ref{lemmaB}}\label{appendix3}
We prove the properties of ${\log \mathcal{V}(Q, \tilde{Q}, \alpha, \beta)}$ under both the HKPI and non-HKPI cases.

\subsubsection{Under HKPI cases}
Firstly, when $Q \in \left( {\beta \tilde Q,\tilde Q} \right)$,  given   $\alpha>$, $0<\beta<1$, and $\beta  < \frac{Q}{{\tilde Q}} < 1$, we have
 \begin{align}
&0 < {B_{\rm{H}}} = \frac{1}{{1 + {e^{\alpha \left( {1 - \beta } \right)}}}} < \frac{1}{2}, \label{eqAppendix1}\\
&0 < {B_{\rm{H}}}\left( {1 + {\cal A}\left( Q \right)} \right) = \frac{{1 + {e^{\alpha \left( {1 - \frac{Q}{{\tilde Q}}} \right)}}}}{{1 + {e^{\alpha \left( {1 - \beta } \right)}}}} < 1.\label{eqAppendix2}
\end{align}
where ${\cal A}\left( Q \right) = {e^{ - \alpha \left( {\frac{Q}{{\tilde Q}} - 1} \right)}}\ge0$.
Then,  the first and second derivatives of ${\log {\cal V}\left( {Q,\tilde Q,\alpha ,\beta } \right)}$ with respect to $\cal Q$ are
 \begin{align}
&\frac{{{\rm{d}}\log {\cal V}\left( {Q,\tilde Q,\alpha ,\beta } \right)}}{{{\rm{d}}Q}} \nonumber\\
 = &\frac{{{\alpha ^2}}}{{\tilde Q}}\left( {\frac{{{B_{\rm{H}}}{\cal A}\left( Q \right)}}{{\left( {1 - {B_{\rm{H}}}\left( {1 + {\cal A}\left( Q \right)} \right)} \right)}} + \frac{{{\cal A}\left( Q \right)}}{{1 + {\cal A}\left( Q \right)}}} \right)\mathop  > \limits_{\left(\rm a \right)} 0,\\
&\frac{{{{\rm{d}}^2}\log {\cal V}\left( {Q,\tilde Q,\alpha ,\beta } \right)}}{{{\rm{d}}{Q^2}}} \nonumber\\
 =&  - \frac{{{\alpha ^3}}}{{{{\tilde Q}^2}}}\left( {\frac{{{B_{\rm{H}}}\left( {1 - {B_{\rm{H}}}} \right){\cal A}\left( Q \right)}}{{{{\left( {1 - {B_{\rm{H}}}\left( {1 + {\cal A}\left( Q \right)} \right)} \right)}^2}}} + \frac{{{ {{\cal A}\left( Q \right)}}}}{{{{\left( {1 + {\cal A}\left( Q \right)} \right)}^2}}}} \right)\mathop  < \limits_{\left( \rm b \right)} 0,
  \end{align}
where   $\rm(a)$ and $\rm(b)$ are due to \eqref{eqAppendix2} and \eqref{eqAppendix1}, respectively.

Moreover, since ${ {\cal V}\left( {Q,\tilde Q,\alpha ,\beta } \right)}=0$ when $Q \in \left( {0, \beta \tilde Q} \right)$ and
 ${ {\cal V}\left( {Q,\tilde Q,\alpha ,\beta } \right)}=1$ when $Q \in \left[ {\tilde Q, +\infty} \right)$,  it holds that
${ \log {\cal V}\left( {Q,\tilde Q,\alpha ,\beta } \right)}$ is nondecreasing over $Q \in (0, +\infty)$ and concave over $Q \in (\beta \tilde{Q},+\infty)$.

\subsubsection{Under LKPI Cases}

Similarly, when  $Q \in \left( {\tilde Q,\frac{{\tilde Q}}{\beta }} \right)$, given  $\alpha>0$, $0<\beta<1$, and $1 < \frac{Q}{{\tilde Q}} < \frac{1}{\beta}$, we have
 \begin{align}
& 0 < {B_{\rm{L}}} = \frac{1}{{1 + {e^{\alpha \left( {\frac{1}{\beta } - 1} \right)}}}} < 1,\label{eqAppendix3}\\
&0 < {B_{\rm{L}}}\left( {1 + {\cal D}\left( Q \right)} \right) = \frac{{1 + {e^{\alpha \left( {\frac{Q}{{\tilde Q}} - 1} \right)}}}}{{1 + {e^{\alpha \left( {\frac{1}{\beta } - 1} \right)}}}} < 1\label{eqAppendix4},
\end{align}
where ${\cal D}\left( Q \right) = {e^{\alpha \left( {\frac{Q}{{\tilde Q}} - 1} \right)}}$.
Then,   the first and second derivatives of ${\log {\cal V}\left( {Q,\tilde Q,\alpha ,\beta } \right)}$ with respect to $  Q$ are
 \begin{align}
&\frac{{{\rm{d}}\log V\left( {Q,\tilde Q,\alpha ,\beta } \right)}}{{{\rm{d}}Q}} \nonumber\\
 =  &- \frac{{{\alpha ^2}}}{{\tilde Q}}\left( {\frac{{{B_{\rm{L}}}{\cal D}\left( Q \right)}}{{\left( {1 - {B_{\rm{L}}}\left( {1 + {\cal D}\left( Q \right)} \right)} \right)}} + \frac{{{\cal D}\left( Q \right)}}{{\left( {1 + {\cal D}\left( Q \right)} \right)}}} \right)\mathop  < \limits_{\left( {\rm{c}} \right)} 0,\\
&\frac{{{{\rm{d}}^2}\log V\left( {Q,\tilde Q,\alpha ,\beta } \right)}}{{{\rm{d}}{Q^2}}}\nonumber\\
 = & - \frac{{{\alpha ^3}}}{{{{\tilde Q}^2}}}\left( {\frac{{{B_{\rm{L}}}\left( {1 - {B_{\rm{L}}}} \right){\cal D}\left( Q \right)}}{{{{\left( {1 - {B_{\rm{L}}}\left( {1 + {\cal D}\left( Q \right)} \right)} \right)}^2}}} + \frac{{{\cal D}\left( Q \right)}}{{{{\left( {1 + {\cal D}\left( Q \right)} \right)}^2}}}} \right)\mathop  < \limits_{\left( {\rm{d}} \right)} 0,
  \end{align}
  where   $\rm(c)$ and $\rm(d)$ are due to \eqref{eqAppendix4} and  \eqref{eqAppendix3}, respectively.

Moreover, since ${ {\cal V}\left( {Q,\tilde Q,\alpha ,\beta } \right)}=1$ when $Q \in \left( {0, { \tilde Q}} \right)$ and
 ${ {\cal V}\left( {Q,\tilde Q,\alpha ,\beta } \right)}=0$ when $Q \in \left[ {\frac{ \tilde Q}{\beta}, +\infty} \right)$,  we proved that
${ \log {\cal V}\left( {Q,\tilde Q,\alpha ,\beta } \right)}$ is nonincreasing over $Q \in (0, +\infty)$ and concave over $Q \in (0, \tilde{Q}/\beta)$.
Finally, Lemma~\ref{lemmaB} is proved.

\subsection{Proof of Theorem~\ref{lemma4}} \label{appendix4}
Firstly, we introduce the following lemma:
\begin{lemma}\label{lemmaBsub}
For a function $h: {\mathbb R} \to {\mathbb R}$ and a function $g: {\mathbb R^{D}} \to {\mathbb R}$, the composition function
$h\left( {g\left( \bf x \right)} \right)$ is concave if one of the following conditions holds:
\begin{itemize}
\item  $h$ is concave and nondecreasing, and $g$ is concave.
 \item $h$ is concave and nonincreasing, and $g$ is convex.
\end{itemize}
\begin{IEEEproof} Please refer to \cite{boyd2004convex} for details.
\end{IEEEproof}
\end{lemma}

From \eqref{eqQC}, \eqref{eq12}, and \eqref{eq21}, it follows that ${V_{kmn}^{{\rm{X}},i}}\ge0$ when ${\bf z}_n \ge  {{\bf{z}}_n^{{\rm{min}}}}$.
Furthermore, by combining Lemmas~\ref{lemmaA}, \ref{lemmaB}, and \ref{lemmaBsub}, we conclude that,
 $\log {V_{kmn}^{\mathrm{C},1}}$,  $\log {V_{kmn}^{\mathrm{P},i}}$ for $1 \le i \le 3$, and $\log {V_{kmn}^{\mathrm{S},1}}$ are non-decreasing and concave functions when ${\bf a}_n$ is given. Finally, Theorem~\ref{lemma4} is proved.

\ifCLASSOPTIONcaptionsoff

\fi
  \bibliography{SPT}
\bibliographystyle{IEEEtran}

\begin{IEEEbiography}[{\includegraphics[width=1in,height=1.25in,clip,keepaspectratio]{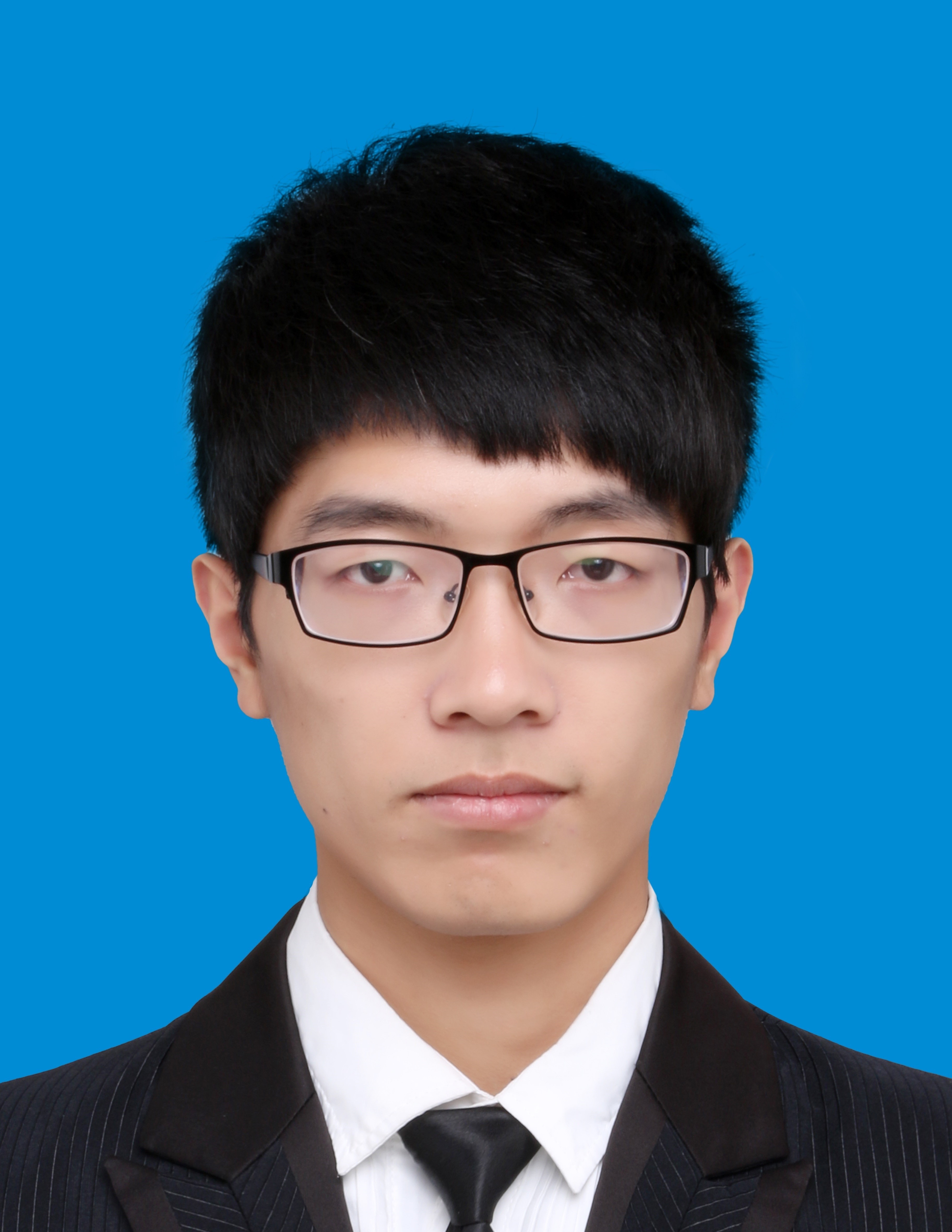}}]{Jie Chen} (Member, IEEE) received the B.S. degree in communication engineering from Chongqing University of Posts and Telecommunications, China, in 2016, and the Ph.D. degree from University of Electronic Science and Technology of China (UESTC), China, in 2021.
From 2019 to 2020, he was a Visiting Student at the University of Toronto, Toronto, ON, Canada.
He is currently a Postdoctoral Research Fellow with the Department of Electrical and Computer Engineering, Western University, London, ON, Canada.
His research interests include integrated sensing and communications, transceiver design for Internet-of-Things, and machine learning for wireless communications.
He was the recipient of the Journal of Communications and Information Networks (JCIN) Best Paper Award in 2021.
\end{IEEEbiography}

\begin{IEEEbiography}[{\includegraphics[width=1in,height=1.25in,clip,keepaspectratio]{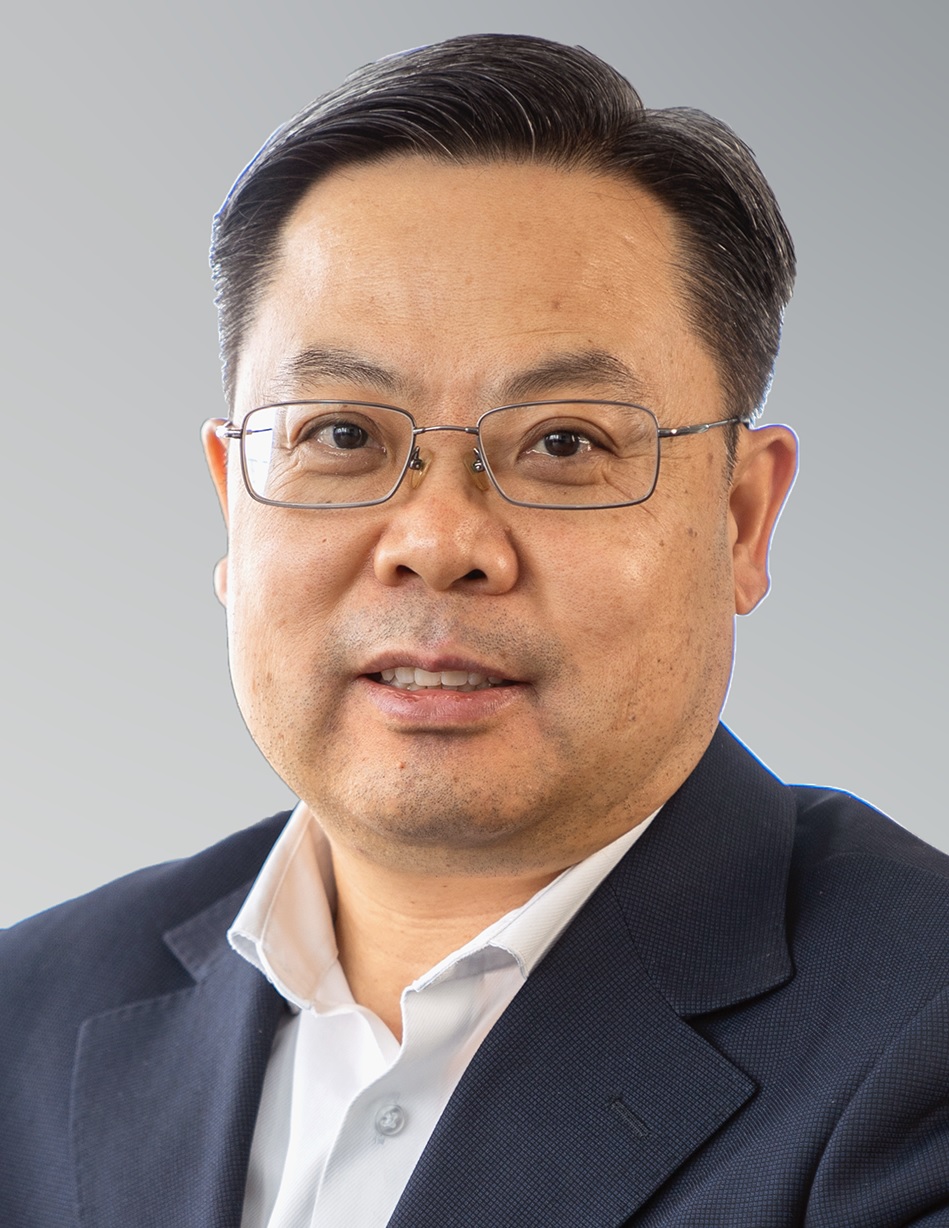}}]{Xianbin Wang} (Fellow, IEEE)
received his Ph.D. degree in electrical and computer engineering from the National University of Singapore in 2001.

He has been with Western University, Canada, since 2008, where he is currently a Distinguished University Professor and Tier-1 Canada Research Chair in Trusted Communications and Computing. Prior to joining Western University, he was with the Communications Research Centre Canada as a Research Scientist and later a Senior Research Scientist from 2002 to 2007. From 2001 to 2002, he was a System Designer at STMicroelectronics. His current research interests include 5G/6G technologies, Internet of Things, machine learning, communications security, digital twin, and intelligent communications. He has over 700 highly cited journals and conference papers, in addition to over 30 granted and pending patents and several standard contributions.

Dr. Wang is a Fellow of the Canadian Academy of Engineering and a Fellow of the Engineering Institute of Canada. He has received many prestigious awards and recognitions, including the IEEE Canada R. A. Fessenden Award, Canada Research Chair, Engineering Research Excellence Award at Western University, Canadian Federal Government Public Service Award, Ontario Early Researcher Award, and twelve Best Paper Awards. He is currently a member of the Senate, Senate Committee on Academic Policy and Senate Committee on University Planning at Western. He also serves on NSERC Discovery Grant Review Panel for Computer Science. He has been involved in many flagship conferences, including IEEE GLOBECOM, ICC, VTC, PIMRC, WCNC, CCECE, and ICNC, in different roles, such as General Chair, TPC Chair, Symposium Chair, Tutorial Instructor, Track Chair, Session Chair, and Keynote Speaker. He was nominated as an IEEE Distinguished Lecturer multiple times by different societies including BTS, ComSoc and VTS. He serves/has served as the Editor-in-Chief, Associate Editor-in-Chief, Area Editor, and editor/associate editor for over ten journals. He was the Chair of the IEEE ComSoc Signal Processing and Computing for Communications (SPCC) Technical Committee and the Central Area Chair of IEEE Canada.
\end{IEEEbiography}

\begin{IEEEbiography}[{\includegraphics[width=1in,height=1.25in,clip,keepaspectratio]{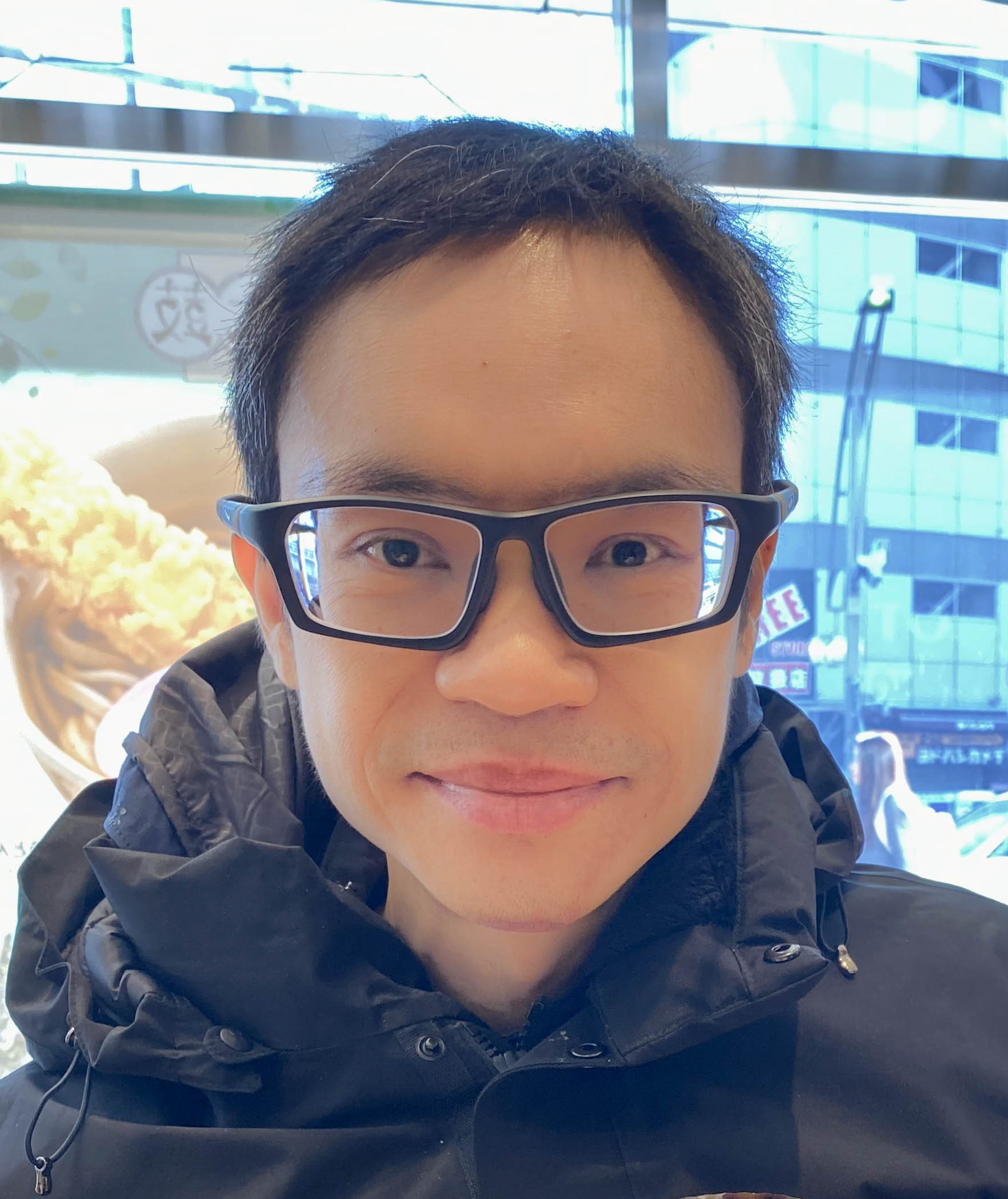}}]{
Dusit Niyato}  (M'09-SM'15-F'17) is a professor in the College of Computing and Data Science, at Nanyang Technological University, Singapore. He received B.Eng. from King Mongkuts Institute of Technology Ladkrabang (KMITL), Thailand and Ph.D. in Electrical and Computer Engineering from the University of Manitoba, Canada. His research interests are in the areas of mobile generative AI, edge general intelligence, quantum computing and networking, and incentive mechanism design.
\end{IEEEbiography}

\end{document}